\documentclass[twocolumn]{aa}  
\usepackage{graphicx}
\usepackage{txfonts}
\usepackage{epsfig}
\usepackage{graphicx}
\usepackage{rotating,lscape}
\begin{document}
   \title{On the Origin of Dwarf Elliptical Galaxies: The Fundamental Plane.}

\author{J. Alfonso L. Aguerri\inst{1} \& A. C\'esar Gonz\'alez-Garc\'\i a\inst{2}}

\offprints{jalfonso@iac.es}

\institute{Instituto de Astrof\'\i sica de Canarias, Calle Via Lactea
  s/n, E-38200 La Laguna, Spain \and
Dept. F\'\i sica Te\'orica, Universidad Aut\'onoma de Madrid, 28049, Madrid, Spain}

   \date{\today}


\authorrunning{Aguerri \& Gonzalez-Garc\'\i a}
\titlerunning{Fundamental Plane of Bright and Dwarf Elliptical Galaxies}


\abstract{Early-type dwarf galaxies are the most common type of galaxies observed in the Universe. Their study has important cosmological implications because according with hierarchical galaxy evolution theories they are the progenitors of brighter galaxies. Nevertheless, the origin of this kind of systems is still not well understood.}{The aim of the present work is 
to investigate whether the different locations of dwarf galaxies with respect to ellipticals in the face-on view of the fundamental plane could be due to the transformation of bright disc galaxies in low-mass systems by harassment.}{We have run high-resolution N-body numerical simulations to test the tidal stripping scenario of the dE galaxies. The present simulations modelled several individual tidal stripping events on initial disc-like galaxy models with different bulge-to-disc mass ratios.}{The models have shown that tidal stripping is a very efficient mechanism for removing stars and dark matter particles from galaxies, specially from their outer parts. The particles of the disc and halo components were easily stripped, while the bulge not. Thus, the scale length of the discs were 40-50$\%$ shorter than the initial ones. Prograde tidal interactions create tidal features like stable bars in the discs of the galaxies. In contrast, bars are inhibited in retrograde encounters. After several tidal interactions the galaxy remnants looks like a dwarf spheroidal system. The final position of the low-mass systems in the face-on view of the fundamental plane depends on the initial conditions of the simulations. Thus, simulated galaxies with initial large B/D ratios are closer to the face-on view of the fundamental plane defined by bright E and bulges of early-type galaxies. Nevertheless, galaxies with initial small B/D ratio are located, after four fast tidal encounters, at the position of dE galaxies in the face-on view of the fundamental plane.The final position of the remnants in the FP do not depend on the orbit configuration of the encounters.}{We conclude that fast galaxy-galaxy interactions are efficient mechanisms for transforming bright galaxies in dwarf ones. Indeed, the different location observed between Es and dEs in the face-on view of the fundamental plane can be explained by the formation of dwarf galaxies by harassment of late-type bright ones.}


   \maketitle
%

\section{Introduction}

Early-type dwarf  galaxies are low luminous ($M_{B} > -18$) and red  systems mostly located in galaxy clusters (\cite{phillipps98}; \cite{pracy04}). The study of these objects has cosmological interest because according with the hierarchical theory of galaxy evolution, dwarfs are the building blocks of the brightest galaxies (\cite{kauffmann93}). Nevertheless, their origin is still matter of debate. Three are the most important formation scenarios of early-type dwarf galaxies: gravitational collapse similar to bright elliptical (E) galaxies, ram-pressure stripping from gas-rich low mass irregular galaxies, and tidal debris left from interactions between galaxies or with the gravitational potential in galaxy clusters.

In the standard picture, early-type dwarf galaxies are formed from the gravitational collapse of primordial density fluctuations, similar to bright E galaxies. Once the first stars are formed, mechanisms of energy feedback into the interstellar medium have been proposed in order to regulate the subsequent star formation or even change the structure of the galaxies (\cite{dekel86}). This standard picture is mainly supported due to several tight correlations between the structural parameters of bright and dwarf early-type galaxies. Those relations, discovered in Virgo, Fornax and Coma clusters, include those between colour and magnitude (\cite{sandage78}), radius and luminosity (\cite{graham03}; \cite{gutierrez04}; \cite{aguerri04}), surface brightness and luminosity (\cite{kormendy77}; \cite{binggeli91}; \cite{aguerri05a}; \cite{graham03}), or S\`ersic shape parameter and luminosity (\cite{graham03}; \cite{aguerri05a}). These tight correlations could indicate a common origin between bright E and early-type dwarf galaxies. However, those common relations have also been explained by some authors as being more due to the uniformity of their dark matter halos than to a common formation mechanism (\cite{navarro97}).

The second formation scenario is based on the presence of common stellar distributions of early-type dwarf and dwarf irregular (dIrr) galaxies (\cite{lin83}; \cite{kormendy85}; \cite{caldwell87}; \cite{binggeli91}; \cite{vanzee04}). The role played by the environment in this formation scenario is crucial. Thus, ram pressure stripping can disrupt the gas from gas rich low-mass galaxies and transform them into  early-type dwarf galaxies (\cite{boselli08}; \cite{sanchezjanssen08}). Repeated tidal shocks suffered by a dwarf satellite galaxy can also remove the kinematic signature of a dIrr galaxy into an early-type dwarf (\cite{mayer01a}; \cite{mayer01b}). This formation mechanism has been reinforced recently by kinematics studies of early-type dwarf galaxies in the Virgo cluster (\cite{pedraz02}; \cite{geha02}; \cite{geha03}; \cite{vanzee04b}; \cite{corsini07}). It was found that most of the early-type dwarf galaxies are rotationally supported similar to dIrrs. But, a scenario where all early-type dwarf galaxies evolve from dIrr via disc fading do not however seem possible, because many of early-type dwarfs  in the Virgo and Fornax clusters are brighter than the dIrr (\cite{conselice01}). Another important point against this scenario is the increase of the metallicity measured in the old stellar populations in early-type dwarf galaxies with respect to dwarf irregulars of similar luminosity (\cite{grebel03}). Although this objection was recently explained by Boselli et al. (2008). 

In the harassment scenario the cumulative effects of tidal fast encounters between galaxies and with the gravitational potential of the galaxy cluster can produce a dramatic morphological transformation from bright spiral galaxies to dwarfs (\cite{moore96}; \cite{mastropietro05}). The effects of the harassment can be seen in the scale length of the discs of bright spiral galaxies. Thus, Aguerri et al. (2004) and Gutierrez et al. (2004) found that the scale length of the discs of bright spiral galaxies of the Coma cluster was shorter than similar galaxies in the field. The disrupted stars could form part of the so-called diffuse light observed in some nearby clusters such as Virgo (\cite{arnaboldi02}; \cite{feldmeier04}; \cite{aguerri05b}; \cite{williams07}), Coma (Gerhard et al. 2005, 2007) and Fornax (\cite{theus97}). However, the major part of this diffuse cluster component seems to be associated with the merger history of the brightest cluster galaxies rather than with disrupted dwarf galaxies (\cite{murante07}). Nevertheless, some bright spiral galaxies can lose part of their stars by fast interactions with the cluster potential and transform into a hot dwarf-like object similar to the observed early-type dwarf galaxies (\cite{mastropietro05}). As previous dwarf formation scenario has also some objections. In particular, one prediction of the harassment model is that dwarf galaxies should lie in low surface brightness tidal streams or arcs. However, Davies et al. (2008) have found no evidence for any such features in a sample of dwarf galaxies in the Virgo cluster.

In a recent series of papers, Lisker et al. (2006a; 2006b; 2007) have proposed that the cluster dwarf galaxies population is composed of different sub-categories of objects, of which not all have been formed in the same way. The different formation mechanisms among the population of early-type dwarf galaxies could be seen in different positions on the fundamental plane (FP). The analysis of the location in the FP of harassed bright spiral galaxies is one piece of information lost in this puzzle, and will be analysed in the present paper.

The FP is a tight relation between the surface brightness, effective radius and central velocity dispersion followed by hot systems such as elliptical galaxies and bulges of spirals. One of the projections of this plane is the so-called Faber-Jackson (\cite{faber76}; FJ) relation. It has been observed that dE and bright E galaxies have different FP (\cite{bender92}; \cite{peterson93}) and Faber-Jackson relations (\cite{held92}; \cite{tonry81}; \cite{davies83}; \cite{matkovic05}). In particular, the face-on view of the FP shows that dE are located in a disjoint area from bright E galaxies. It was also observed that bright E galaxies follow the FJ relation which can be expressed as $L \propto \sigma^{n}$, where $n\approx4$ (\cite{faber76}; \cite{sargent77}; \cite{schechter79}; \cite{schechter80}; \cite{tonry81}; \cite{terlevich81}; \cite{forbes99}), while dE galaxies show $n\approx 2$ (\cite{tonry81}; \cite{davies83}; \cite{held92}; \cite{matkovic05}). These observational evidences suggest a different origin and/or evolution of these two kinds of systems. Thus, Matkovi\'c \& Guzm\'an (2005) demonstrated that the different FJ relation observed between bright E and dE can not be explained with the presence of rotation in the dE sample. They conclude that faint early-type galaxies in Coma had to be formed at least 6 Gyr ago and over a short 1 Gyr period. The different FP displayed by bright E and dE was also investigated by de Rijcke et al. (2005). They pointed out that internal processes could produce those differences. They concluded that semi-analytical models that incorporate quiescent star formation combined with post-merger starbursts and the dynamical response after supernova-driven gas-loss, were able to reproduce the position of dE in the FP. These two previous works indicate that the differences in the FP and FJ relation could be explained by different internal evolution between bright E and dE galaxies. Nevertheless, few attention has been given until now to the analysis of these differences due to other origins.

The origin and evolution of the FP of elliptical galaxies has received considerable attention from N-body modellers (\cite{capelato95, bekki98, aceves05}). Gonz\'alez-Garc\'\i a \& van Albada (2003) and Nipoti et al. (2003) showed that the FP is not severely affected by a few merging events. This results have been confirmed by several latter studies (e.g. Boylan-Kolchin, Ma \& Quataert 2006) including self-consistent cosmological simulations (O\~norbe et al. 2006, 2007). However the location of the different galaxy types, like dEs, in the FP has received much less interest.

The aim of the present work is to investigate whether the position of the dE in the FP can be explained by the harassment scenario, which posits that dEs stem from perturbed and stripped late-type disc galaxies that entered in clusters and groups of galaxies some Gyr ago. The structure of the paper is as follow: the sample of galaxies used and the FP of bright and low-luminous galaxies are shown in Section 2; N-body numerical simulations of harassed bright  galaxies are presented in Section 3; our discussion and conclusions are given in Sections 4 and 5, respectively.

\section{The Fundamental Plane \label{sec:FP}}

\begin{figure}
\begin{center}
\mbox{\epsfig{file=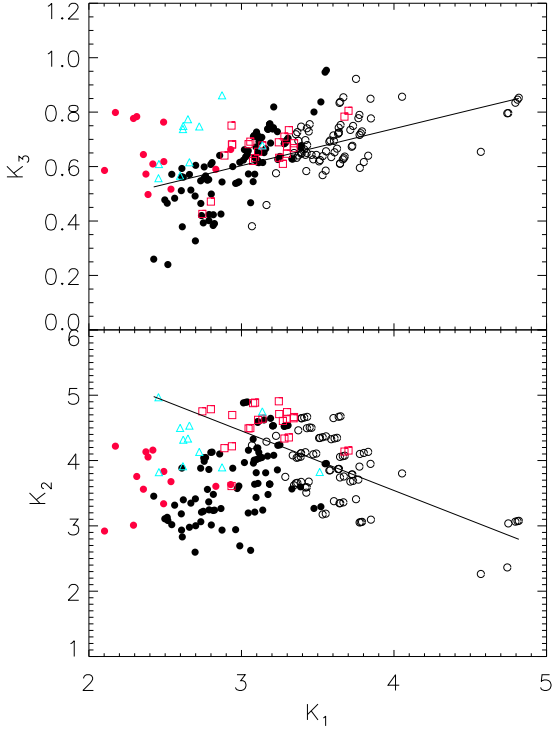,width=8cm}}
\end{center}
\caption{Fundamental plane in the $\kappa$ representation of the different types of galaxies in the Coma cluster: bright E (empty circles), dE (black circles), early-type bulges (red squares), late-type bulges (blue triangles), and dS0 bulges (red circles). The full lines shows the fits of the FP in the $\kappa$-space defined only by bright Es.}
\label{Fig:fig1}
\end{figure}

\begin{figure}
\begin{center}
\mbox{\epsfig{file=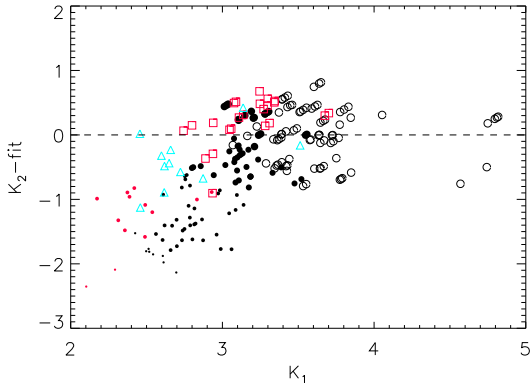,width=8cm}}
\end{center}
\caption{Residuals of the face-on fundamental plane in the $\kappa$ representation of the different types of galaxies in the Coma cluster: bright E (empty circles), dE (black circles), early-type bulges (red squares), late-type bulges (blue triangles), and dS0 bulges (red circles).}
\label{Fig:fig2}
\end{figure}

It is well known that dynamically hot systems (elliptical galaxies and bulges of spiral galaxies) are not scattered randomly in the three-dimensional space spanned by the luminosity, half-light radius, and velocity dispersion (Dressler et al. 1987; Djorgovski \& Davis 1987; Bender, Burstein \& Faber 1992; Falc\'on-Barroso, Peletier \& Balcells 2002; Bernardi et al. 2003) . Those objects are located in a plane called the Fundamental Plane given by:

\begin{equation}
\log(r_{e})=\alpha \log(\sigma_{\rm o}) + \beta \langle \mu_{e} \rangle + \gamma
\end{equation}
where $\alpha$, $\beta$, and $\gamma$ are constants, and $r_{e}$, $\langle \mu_{e} \rangle$, and $\sigma_{\rm o}$ are the half-light effective radius, mean surface brightness and central velocity dispersion of the galaxies, respectively.

The FP can be directly derived from the virial theorem:

\begin{equation}
\langle v^{2} \rangle=GM/r_{g}
\end{equation}
where $M$ is the total mass of the system, $r_{g}$ is known as the gravitational radius and $\langle v^2 \rangle$ is the mean square velocity of the stars in the system.  To relate these quantities to observational ones, Capelato et al. (1995) introduced:

\begin{equation}
C_{v}\equiv \langle v^{2} \rangle/\sigma_{\rm o}^{2}\\  
C_{r} \equiv r_{g}/r_{e}
\end{equation}

Introducing these parameters in eq (1), and using $\langle \mu_{e} \rangle=-2.5 \log( \frac{L}{2\pi r_{e}^{2}})$, being $L$ the luminosity of the object, the FP can be written as:

\begin{equation}
\log(r_{e})=2 \log(\sigma_{\rm o})+ 0.4 \langle \mu_{e} \rangle + \log(C_{v}C_{r})-\log(\frac{M}{L})+\gamma^{'}
\end{equation}

Thus, the virial theorem predicts that $\alpha=2$ and $\beta=0.4$, while the observed values are $\alpha=1.24$ and $\beta=0.328$ (J\o rgensen, Franx \& Kj\ae rgaard, 1996). Equations (1) and (4) determine two oblique planes defining an angle, the so-called tilt of the FP. From eq. (4) can be seen that the tilt of the FP is due to differences in the structure or in the M/L ratio of the galaxies.

Another formulation of the FP is the so-called $\kappa$-space (Bender et al. 1992) expressed by:

\begin{equation}
\kappa_{1}=\frac{1}{\sqrt{2}}\log(r_{e}\sigma_{\rm o}^{2}); \\
\kappa_{2}=\frac{1}{\sqrt{6}}\log(\frac{I_{e}\sigma_{\rm o}^{2}}{r_{e}}); \\
\kappa_{3}=\frac{1}{\sqrt{3}}\log(\frac{\sigma_{\rm o}^{2}}{I_{e}r_{e}})
\end{equation}

Using these definitions and the virial theorem, $\kappa_{1}$ is expected to be a measure for the mass of the galaxy, $\kappa_{2}$ is mostly sensitive to surface brightness, and $\kappa_{3}$ depends on the mass-to-light ratio. In this parameterisation, the $\kappa_{1}-\kappa_{3}$ plane is an edge-on view of the FP, while the $\kappa_{1}-\kappa_{2}$ plane gives the face-on view of the FP.

Most of the studies about the FP have been centred on bright early-type galaxies in nearby galaxy clusters (see D'onofrio et al. 2006, and references therein). Notwithstanding, the location of dwarf galaxies in the fundamental plane has not been extensively analysed after the early works by Nieto et al. (1990), Bender et al. (1992), and Guzman et al. (1993). However, Bender et al. (1992) and Guzman et al. (1993), based on a small sample of dwarf galaxies, found that dwarf galaxies were located in a different position in the face-on view of the FP defined in the $\kappa$-space. The distribution of the dwarf galaxies were perpendicular to the relationship defined by luminous elliptical galaxies in the $\kappa_{1}-\kappa_{2}$ plane. This trend was also recently confirmed on a larger sample by De Rijcke et al. (2005). 

In the present work we have analysed the fundamental plane of a large sample of bright and dwarf galaxies located in the Coma cluster. The sample is formed by 215 galaxies of different types: 35 bright E, 113 dE, 12 dS0, 23 bulges of early-type spirals, and 11 bulges of late-type galaxies\footnote{See Aguerri et al. (2004, 2005a) for details on the galaxy classification}. The photometric parameters of the galaxies (r$_{e}$ and $\mu_{e}$) were obtained from the surface brightness decompositions on these objects carried out by Aguerri et al. (2004, 2005a), and the central velocity dispersion of the galaxies ($\sigma_{o}$) was taken from Matkovic \& Guzm\'an (2005) and Smith et al. (2004).

Figure 1 shows the edge-on and face-on representations of the FP in the $\kappa$-space for the galaxies in the Coma cluster.  We have fitted, in both orientations, the FP determined only by bright E galaxies. The scatter around the edge-on view of this FP depends on the galaxy type, being: 0.07, 0.11, 0.11, 0.07, 0.23 for bright E, dE, dS0, early-type bulges and late-type bulges, respectively. Notice that low-luminous galaxies and late-type bulges have larger dispersion than bright E and early-type bulges. This is in agreement with previous findings  by Nieto et al. (1990). The segregation between low-mass systems and bright E galaxies is clear in the $\kappa_{1}-\kappa_{2}$ plane (face-on view of the FP; see Fig. 1). As it was observed in previous works (Bender et al. 1992; Guzman et al. 1993; De Rijcke et al. 2005), dwarf galaxies and late-type bulges have different location on the face-on view of the FP than bright E and early-type bulges. This difference is more clear in Fig. 2, which shows the residual of each galaxy respect to the fitted face-on view of the FP defined by bright E galaxies. In Fig. 2 we have scaled the symbols of the dwarf galaxies by their $\sigma_{o}$ value. Notice that dwarf galaxies with low values of $\sigma_{o}$ are the galaxies located further away from the FP defined by bright E.  

The aim of the present work is to understand if the segregation between bright and dwarf galaxies in the face-on view of the FP could be explained by the transformation of bright disc galaxies in low-mass systems by fast galaxy-galaxy encounters.  

\begin{figure}
\begin{center}
\mbox{\epsfig{file=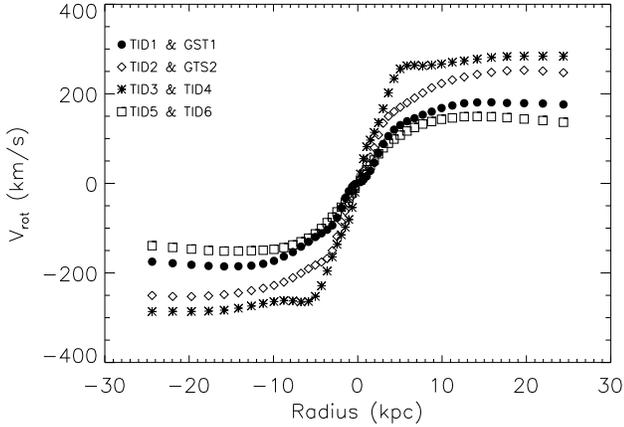,angle=0,scale=0.5}}
\end{center}
\caption{Circular velocity curves for the initial conditions of the different simulations.}
\label{Fig:fig2}
\end{figure}

\section{Numerical Simulations}

We have performed several N-body simulations to test the tidal stripping formation scenario for dEs. The present simulations are not cosmological ones. They model several individual tidal stripping events happening on different initial disc-like galaxy models. The simulations are pure N-body models with no gas content. Nevertheless, we have also simulated the effect of gas stripping on two models by removing 10$\%$ of the luminous particles at the beginning of the simulations.  After each tidal interaction or gas stripping event we allow to the system to reach virial equilibrium and then we have investigated its position in the $\kappa$-space.

\subsection{Initial conditions and units of the simulations}

\begin{table}
\begin{center}
\caption{Input number of particles and mass fraction of each component of the different models. Particle number is given in K-particles ($\times 10^3$).\label{tab1}}
\begin{tabular}{@{}ccccccc}
\hline
{\bf $Model$}&{\bf $\#$}& {\bf $\#_{halo}$} & {\bf $\#_{disc}$} & {\bf $\#_{bulge}$} & {\bf $\frac{M_{lum}}{M_{total}}$} & {\bf $\frac{M_{bulge}}{M_{disc}}$}\\
\hline
TID1&  370 & 300 & 60 & 10 &1/8 & 1/12\\
TID2&  540 & 275 & 175& 90 &1/7.5 & 1/5\\
TID3&  540 & 275 & 175& 90 &1/5 & 1/2\\
TID4&  540 & 275 & 175& 90 &1/5 & 1/2\\
TID5&  400 & 310 &  90& -- &1/6 & --\\
TID6&  400 & 310 &  90& -- &1/6 & --\\
TID7&  400 & 310 &  90& -- &1/6 & --\\
TID8&  400 & 310 &  90& -- &1/6 & --\\
\hline
GST1&  370 & 300 & 60 & 10 &1/8 & 1/12\\
GST2&  540 & 275 & 175& 90 &1/7.5 & 1/5\\
\hline  
\end{tabular}
\end{center}
\end{table}

The initial galaxies of our simulations consisted on disc galaxies generated by the Kuijken \& Dubinsky (1995) disc-bulge-halo algorithm. The galaxy has a King (1966) model for the bulge (if present), a Shu (1969) for the disc and an Evans (Kuijken \& Dubinsky 1994) model for the halo, being all components modelled live. Table \ref{tab1} gives the details of the different initial models of the simulations.

We have simulated galaxies with different bulge-to-disc (B/D) mass ratios in order to test different starting points for the simulations. Thus, the initial B/D mass ratios were 1:12 (TID1 and GST1 models), 1:5 (TID2 and GST2 models), and 1:2 (TID3 and TID4 models). In addition, four more models (TID5, TID6, TID7 and TID8) were pure disc galaxies, where no bulge was considered in the initial conditions. Assuming the well-know observational correlation between morphological type and B/D (e.g. \cite{simien86, aguerri04, mendezabreu08}) we can say that TID3 and TID4 could represent an early-type Sa galaxy; TID2 and GST2 show an Sb galaxy; TID1 and GST1 are similar to Sc galaxies; and  TID5-TID8 can be compared with an Sd/Sm galaxy. 

We used model units to run the simulations. This units take the total mass of the initial galaxy as unity, the disc scale length as 0.2, and the Constant of Gravity G=1. However, in order to compare with real galaxies we need to convert the model units to physical ones. This was made by defining a set of conversion units to compare our simulations with observations. For bright galaxies, the scale length of the disc is independent of the morphological type, being $h\approx3$ kpc (see e.g. \cite{aguerri04} and references therein). This means that we can take a conversion factor of $[L]=15$ kpc for all simulations as the unit of length. In contrast, different galaxy types show different rotation curves. Thus, the maximum velocity ($V_{max}$) of Sa, Sb, Sc, and Sd galaxies are $\sim$300, $\sim$250, $\sim$180 and $\sim$150 km/s, respectively (Roberts 1978; Rubin et al. 1985). This means that the velocity conversion factor for our simulations were $[V]= 496$ km/s for TID3 and TID4; $[V]=345$ km/s for TID2 and GST2; $[V]=230$ km/s for TID1 and GST1; and  $[V]=175$ km/s for TID5-TID8. Figure 3 shows the initial rotation curves of the different simulations. Notice the different $V_{max}$ and rotation curve central gradients of the simulations. Thus, TID5-TID8 models shows the smaller $V_{max}$ and the less steep central gradient in the rotation curve as it is observed in late-type galaxies. In contrast, TID3 and TID4 models show the largest $V_{max}$ and steeper rotation curve as correspond to Sa galaxies with large central mass concentration (large bulges).  Assuming a time unit $[T]=2.4\times10^{7}$ yr, the total mass of the galaxies turned to be: $[M]=8.58\times10^{11} M_{\odot}$ for TID3 and TID4; $[M]=4.15\times10^{11} M_{\odot}$ for TID2 and GST2; $[M]=1.84\times10^{11} M_{\odot}$ for TID1 and GST1; and $[M]=1.07\times10^{11} M_{\odot}$ for TID5-TID8. Thus, in agreement with observations, the more massive galaxies correspond to those models simulating early-type galaxies.

\begin{figure*}
\begin{center}
\mbox{\epsfig{file=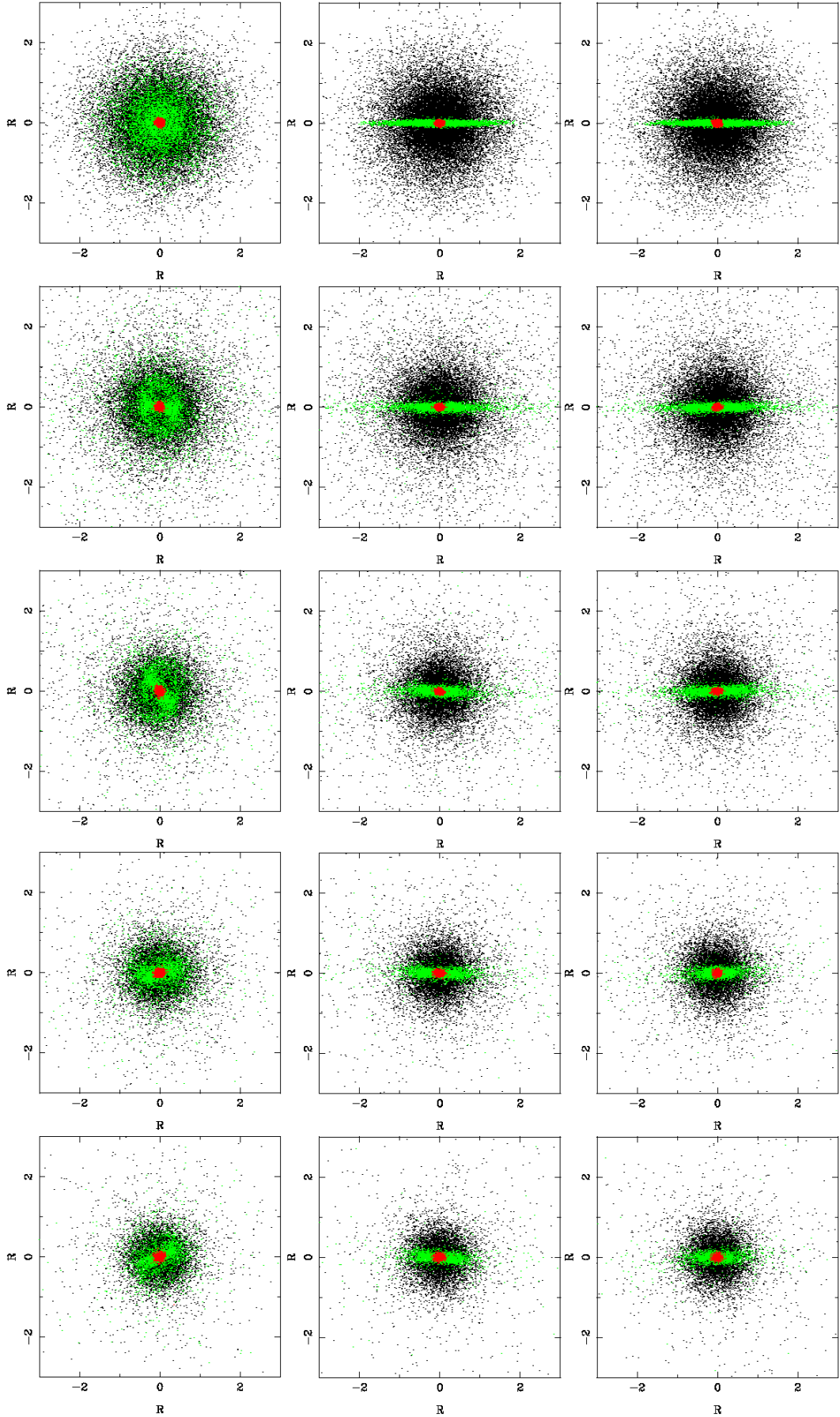,width=14cm}}
\end{center}
\caption{Evolution of the tidal events in TID1. Rows give the three views of the system after the different tidal harassment events. Colours indicate the different constituents of the initial galaxy: black-dark halo, green-disc and red-bulge}
\label{Fig:fig3}
\end{figure*}

\begin{figure*}
\begin{center}
\mbox{\epsfig{file=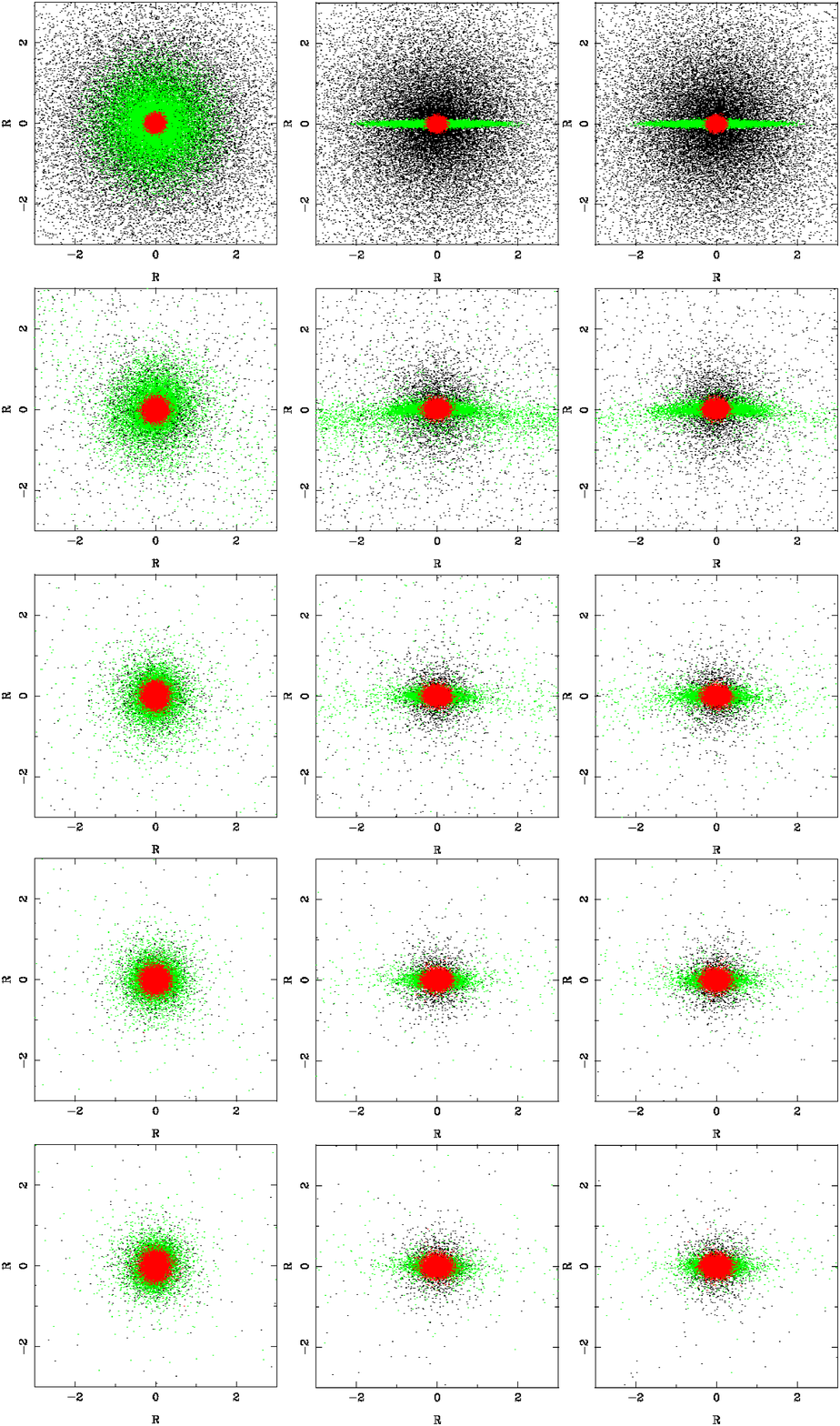,width=14cm}}
\end{center}
\caption{Evolution of the tidal events in TID3. Rows give the three views of the system after the different tidal harassment events. Colours indicate the different constituents of the initial galaxy: black-dark halo, green-disc and red-bulge}
\label{Fig:fig4}
\end{figure*}

\subsection{Galaxy-Galaxy encounters}

The aim of the present simulations is to model fast encounters of our initial disc galaxy with another system of similar mass. Those galactic interactions would be similar to the fast galaxy-galaxy encounters predicted in high density environments as galaxy clusters (Moore et al. 1996). This kind of encounters has the peculiarity that the effective duration of the encounter is shorter than the characteristic crossing time of the majority of their stars. Moreover, the impact parameter of our fast encounters will be much larger than the typical galaxy scales. Therefore, this kind of galaxy encounters can be modelled using the impulse tidal approximation (see Binney \& Tremaine 1987, \S7.2(d)). Thus, we can assume that the galaxies have no time to reorganise during the encounter, hence the energy injection is all kinetic; reorganisation occurs after the encounter has finished.
 
The commonly used formula for the tidal impulse approximation does not take into account the extended nature of both bodies. Nevertheless, our encounters could not be considered between point-like galaxies. To take this into account, we have applied the formalism proposed by  Gnedin, Hernquist \& Ostriker (1999) giving that the variation in the velocity ($\Delta {\bf V_2}$), angular momentum ($\Delta J_2$) and energy ($\Delta E_2$) of the particles in the models are:

\begin{equation}
\mid \Delta {\bf V}_2(R)\mid \sim 2 \frac{G M_{1}}{b^{2}V} \frac{R_{peri}}{R_{max}}R^2 
\label{eqn:Deltav2}
\end{equation}
\begin{equation}
\Delta E_2 (R) \sim  \frac{G^2 M^2_{1}}{b^{4}V^2} R^2 
\label{eqn:Deltae}
\end{equation}
\begin{equation}
\Delta J_2(R) \sim 2 \frac{G M_{1}}{b^{2}V} I_{22}(R), 
\label{eqn:DeltaJ}
\end{equation}

\noindent where $M_1$ is the total mass of the perturber, b is the impact parameter of the orbit, $R_{peri}$ is the pericenter distance, $V$ is the relative velocity at pericenter, $R_{max}$ is the cut-off imposed in our initial model for the perturber galaxy, and $I_{22}$ is the smallest eigenvalue of the
inertia tensor. A detailed discussion of these formulae and their implications can be found in Gonz\'alez-Garc\'\i a, Aguerri \& Balcells (2005).

The fast encounters were simulated injecting energy, according with the previous expressions, to each particle of the modelled galaxy. Stars and dark matter particles in the modelled galaxy absorb energy ($\Delta E \approx 0.5 (\Delta V_{2})^{2}$) as well as angular momentum and may become unbound after the interaction. Since the velocity and energy change depend on the distance to the central parts, the particles with higher binding energy ($E_{bind}$) are the ones that are going to be easily stripped away. After each tidal interaction we allow the system to reach virialization in the innermost regions by evolving the galaxy in isolation for more than 25 time units. This amounts for more than 50 half-mass radius crossing times of the bulge.  For the different runs we have taken $b=\sqrt{2}$, $V=2$, $R_{peri} = R_e$ and $M_1$ equal to the total mass of the initial galaxy at each tidal interaction, so we model equal-mass encounters. A total number of 4 tidal interactions were simulated for each model.

We have run 10 different simulations (see Tab. 1). The models called TID1, TID2, TID3 and TID5 differ on the bulge-to-disc ratio (B/D). Thus, they have $B/D=1/12, 1/5$, $1/2$, and $0$ respectively. The relative orientation of the initial disc with respect to the orbital angular momentum was $(\theta,\phi)=(0,0)$. Models TID4 and TID6 were replicas of TID3 and TID5, changing the orientation of the initial discs with respect to the assumed orbit, now being $(60,30)$. Models TID7 and TID8 are also identical to model TID5 but with retrograde orbits with respect to the disc spin; TID7 has $(\theta,\phi)=(180,0)$, while TID8 has a less symmetric configuration $(\theta,\phi)=(135,0)$. The tidal features of these systems were somewhat different from TID3 and TID5, but the amount of material stripped and the general structure of the remnants were basically the same.  We have also run two more models (GST1 and GST2) simulating an initial gas stripping. Those simulations have identical initial conditions to TID1 and TID2, respectively. The initial gas stripping was simulated by randomly removing 10$\%$ of the disc particles of the models. Gas-stripping depends also on the distance from the centre of the galaxy (see the hydrodynamical simulations by Quilis et al. 2000), but here we adopt an extreme situation where at a single go all gas in the galaxy is stripped away. After they reach virialization we applied the first tidal encounter.

Simulations were run using the {\small GADGET2} (Springel 2005) hydrodynamical code, in the tree-code configuration, on the Beowulf cluster of the Instituto de Astrof\'{\i}sica de Canarias (IAC), the Mare Nostrum supercomputer and the La Palma node of the National Supercomputing Center (BSC). A typical run with $5 \times 10^5$ particles took of the order of $5 \times 10^4$ s in 16 CPUs. We used different softening parameters for the different particles, typically $\epsilon \sim 1/10$ of the half-mass radius of each system. An opening parameter of $\theta = 0.8$ and quadrupole terms were included in the force calculation. 

\subsection{Results}

\begin{table}
\begin{center}
\caption{Half mass radius for each of the components for the initial and final models of our simulations.\label{tab2}}
\begin{tabular}{@{}ccccc}
\hline
{\bf $Model$}&& {\bf $r_{e,halo}$} & {\bf $r_{e,disc}$} & {\bf $r_{e,bulge}$} \\
\hline
TID1&  i& 0.834 & 0.610 & 0.073\\
    &  f& 0.629& 0.403 & 0.075\\
\hline
TID2&  i& 1.050 & 0.592 & 0.071\\
    &  f& 0.681& 0.389 & 0.110\\
\hline
TID3&  i& 1.920 & 0.639 & 0.096\\
    &  f& 0.528& 0.380 & 0.200\\
\hline
TID4&  i& 1.920 & 0.639 & 0.096\\
    &  f& 0.526& 0.365 & 0.199\\
\hline
TID5&  i& 0.629 & 0.731 & ----\\
    &  f& 0.548& 0.427 & ----\\
\hline
TID6&  i& 0.629 & 0.731 & ----\\
    &  f& 0.542& 0.344 & ----\\
\hline
TID7&  i& 0.629 & 0.731 & ----\\
    &  f& 0.521& 0.373 & ----\\
\hline
TID8&  i& 0.629 & 0.731 & ----\\
    &  f& 0.548& 0.412 & ----\\
\hline
GST1&  i& 0.826 & 0.608 & 0.067\\
    &  f& 0.583 & 0.378 & 0.104\\
\hline
GST2&  i& 1.070 & 0.611 & 0.115\\
    &  f& 0.612& 0.374 & 0.130\\
\hline  
\end{tabular}
\end{center}
\end{table}

\begin{figure}
\begin{center}
\mbox{\epsfig{file=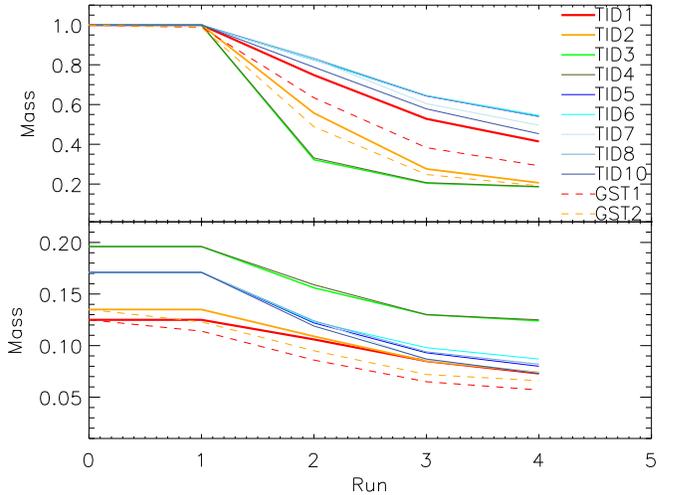,width=9cm}}
\end{center}
\caption{Change of mass in each tidal harassment event. Upper panel:total mass, lower panel: luminous mass.}
\label{Fig:fig7}
\end{figure}

Figures~\ref{Fig:fig3} and ~\ref{Fig:fig4} present the initial and evolved states after each tidal interaction for the TID1 and TID3 simulations, respectively. These two models represent two extreme cases because they have the largest (TID3) and smallest (TID1) initial bulges. In both cases, tidal stripping removes first the outer parts of the halo, after two tidal interactions a fair amount of disc particles are also expelled. Models with large bulges (TID2, TID3) show that almost the whole disc was removed after 4 tidal interactions. These models have at the end a disc-to-bulge ratio close to 1:1, leaving the small number of still bounded disc particles in a roundish component (see Fig.~\ref{Fig:fig4}). In contrast, the model with the smallest initial bulge (TID1) still presents a short and thick disc after 4 tidal encounters (see Fig.~\ref{Fig:fig3}). These truncations have been quantified measuring the half mass radius of the bulge ($r_{e,bulge}$), disc ($r_{e,disc}$) and halo ($r_{e,halo}$) as showed in Table 2. Notice that while the disc and halo are strongly truncated in all simulations, the bulge enlarge its half mass radius specially for the models with the largest B/D ratio (TID3 and TID4).

\begin{figure*}
\begin{center}
\mbox{\epsfig{file=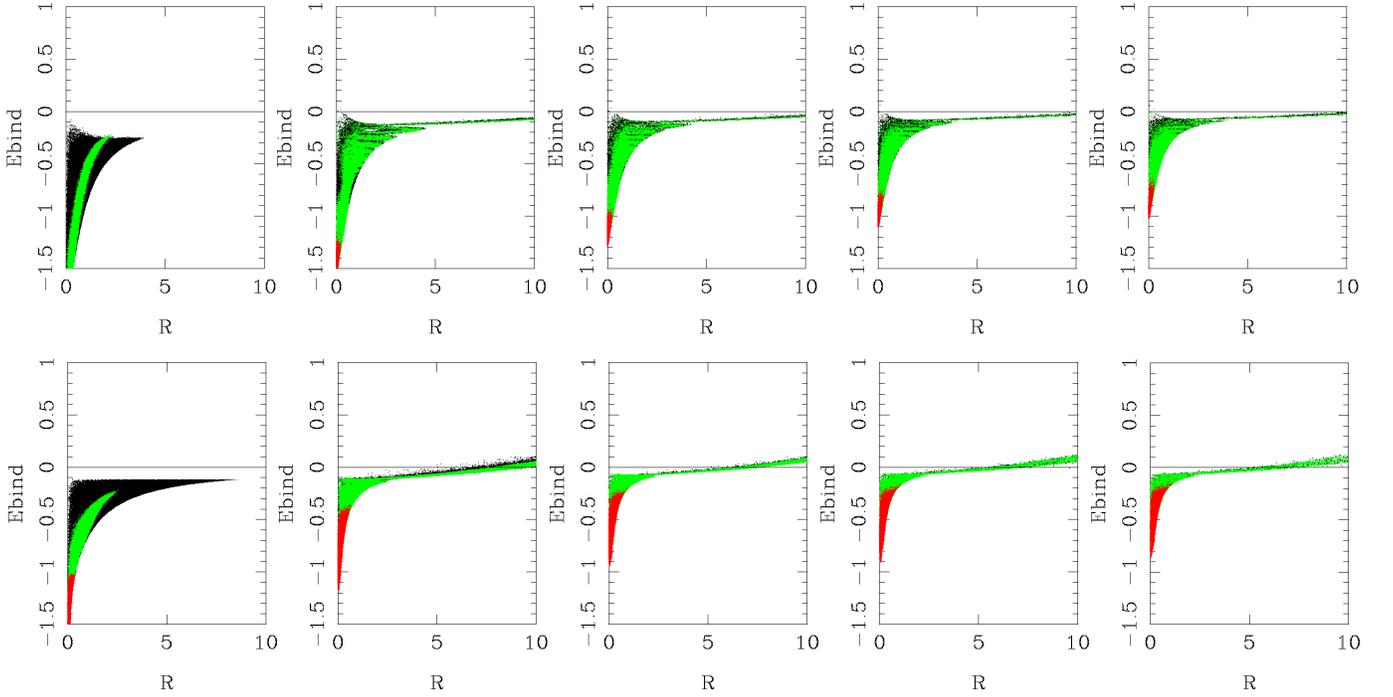,width=18cm}}
\end{center}
\caption{Binding energy vs distance from the centre for the final snapshots of the simulations TID1 (up) and TID3 (bottom). The first panel in each case presents the state of the initial galaxy model.}
\label{Fig:fig6}
\end{figure*}

\begin{figure*}[!htb]
\begin{center}
\mbox{\epsfig{file=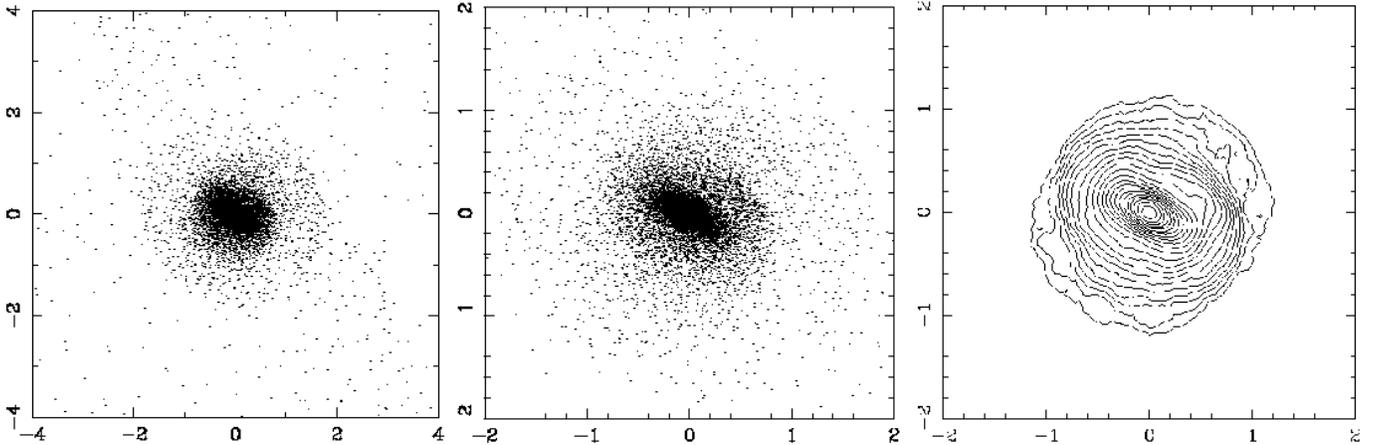,width=18cm}}
\end{center}
\caption{Simulation TID2, snapshot after the evolution of the second tidal harassment event in simulation TID2. Left and central panel show the luminous particles (bulge+disc). We observe the formation of tidal tails (left) and a prominent bar (centre). This bar can clearly be seen in the contour plot (right).}
\label{Fig:fig5}
\end{figure*}

\subsubsection{Disc and Halo truncations}
The truncation of the disc and halo component results in an important mass loss of the galaxies. Figure~\ref{Fig:fig7} presents the evolution of the total content of bound mass in these systems. We find that harassment is highly efficient in removing a large amount of the initial mass in all our simulations. The different encounters removes between $\sim50\%$ to 80$\%$ of the total initial mass, and between 30$\%$ and 60$\%$ of the luminous mass.

\begin{figure*}[|htb]
\begin{center}
\mbox{\epsfig{file=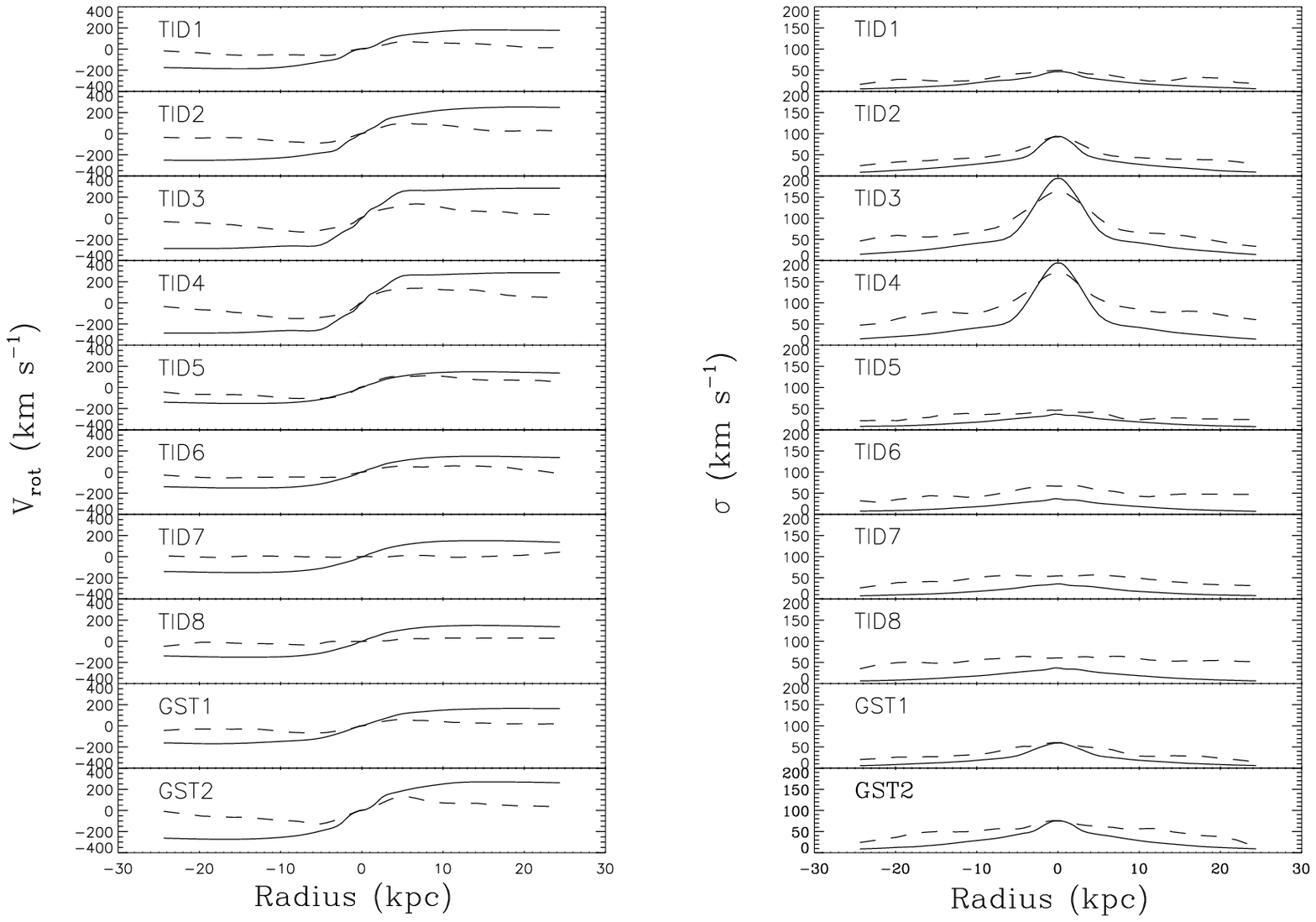,angle=0}}
\end{center}
\caption{Rotation curves (left panels) and dispersion velocity profiles (right panels) for the initial (full lines) and remnants (dashed lines) of the simulations.}
\label{Fig:fig20}
\end{figure*}

Figure~\ref{Fig:fig6} shows the binding energy vs. the radial distance to the galaxy centre after each tidal encounter for TID1 and TID3 models. It is clear that the outermost particles are stripped first after each tidal interaction because they have larger binding energies. Thus, disc and halo particles were easily stripped away contrary to bulge particles. In both models shown in Fig.~\ref{Fig:fig6} the simulated galaxy-galaxy encounters do not provide enough energy to extract the bulge particles. But, some bulge particles occupy larger binding energies previously occupied by striped disc and halo particles producing an inflation of the bulge. This behaviour is common to all the  simulations, being more prominent in those with large B/D-ratio. According with this, we can say that the bulge survive in those simulations with initial bulge. Moreover, at the end of the simulations, the hot component of galaxies with initial bulges is basically formed by the initial bulge of the galaxies.

\subsubsection{Tidally induced bars}
The stripping remnants show a number of tidal features like extended tidal tails, rings, etc, that indicate that the system has suffered such interactions.  

Some systems appear to present a bar-like structure in the disc (see Figure~\ref{Fig:fig5}). This bar-like structures induced by tidal interactions are formed in the first tidal encounter and are stable until the end of the simulation. The bars are more prominent in those models with lower central mass concentration, i.e. smaller bulge component, and a prograde rotation. The bar formation depends on the orbit configuration of the encounters. Thus, retrograde orbits seem to inhibit bar formation. Only the models with the largest bulge component (TID3 and TID4) do not show any bar feature in the disc. This is in agreement with numerical simulations about bar formation and evolution which find that central mass concentrations inhibits bar formation in discs. (Pfenniger \& Norman 1990; Shen \& Sellwood 2004; Athanassoula et al. 2005). 

\begin{figure}
\begin{center}
\mbox{\epsfig{file=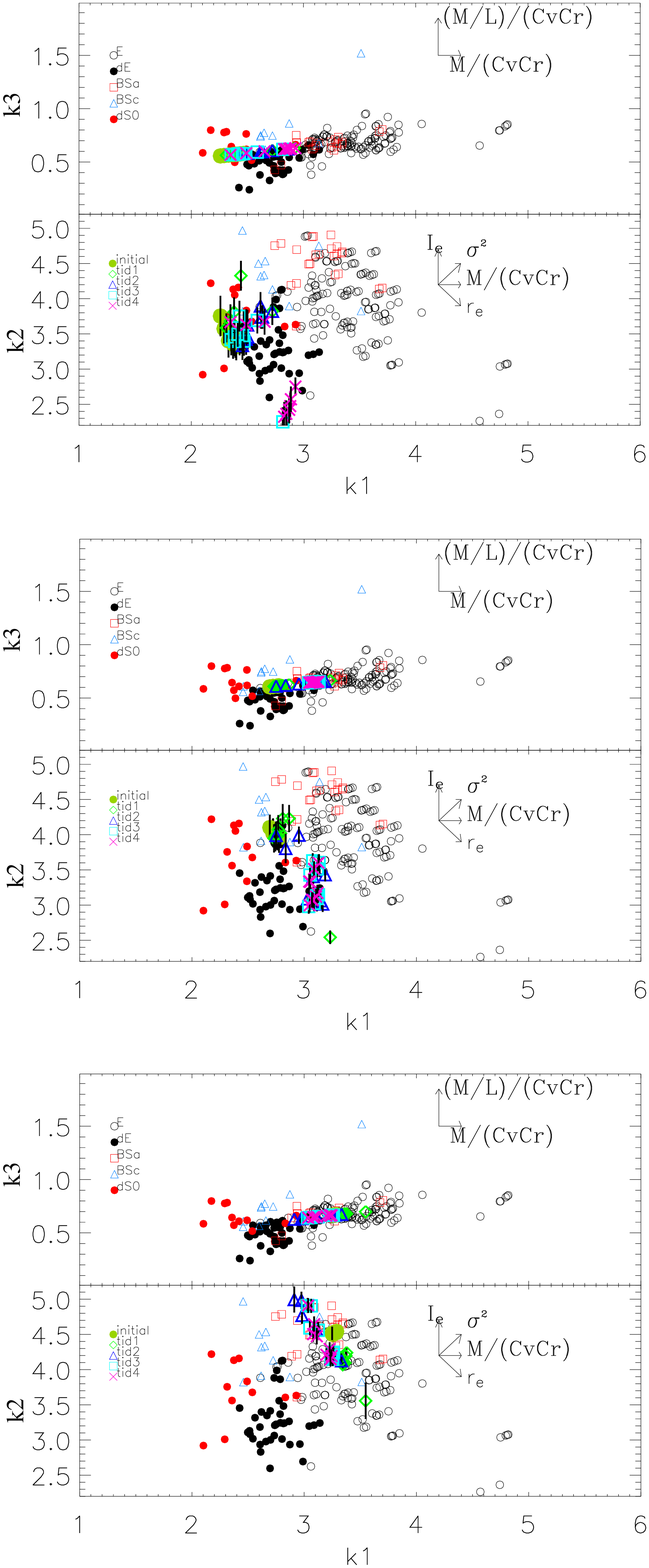,width=8.5cm}}
\end{center}
\caption{Results for the models TID1 (top), TID2 (middle) and TID3 (bottom) presented in the $\kappa$-space. The different types of galaxies in the Coma cluster are indicated as black small symbols: bright E (diamonds), dE (black circles), early-type bulges (red squares), late-type bulges (blue squares), and dS0 bulges (red circles)). Colour symbols indicate the results for the models.}
\label{Fig:fig9}
\end{figure}

\begin{figure}
\begin{center}
\mbox{\epsfig{file=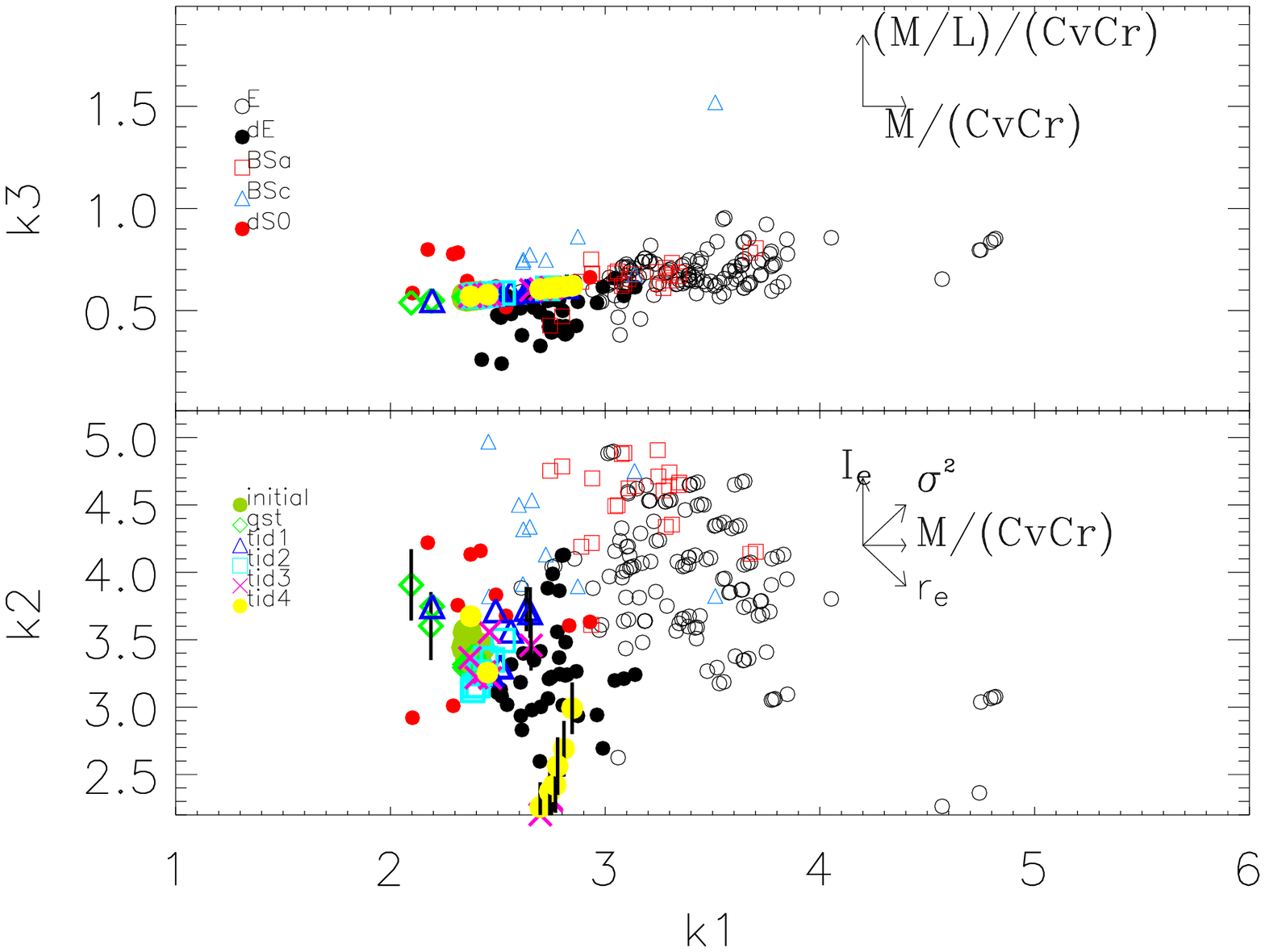,width=8cm}}
\mbox{\epsfig{file=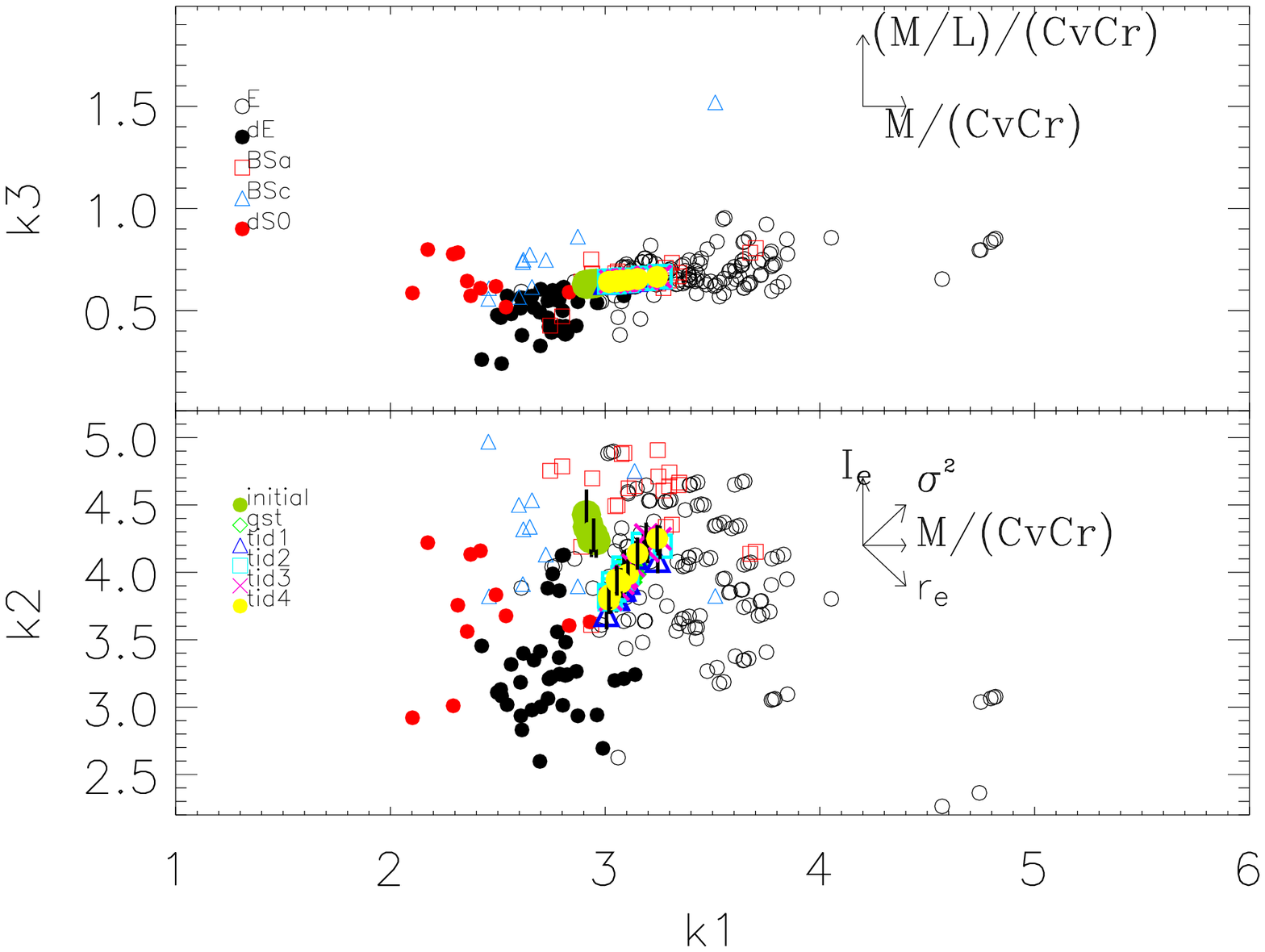,width=8cm}}
\end{center}
\caption{Results for the models GST1 and GST2 presented in the $\kappa$-space. Symbols and colours are as in Figure\ref{Fig:fig2}. }
\label{Fig:fig10}
\end{figure}

\begin{figure}
\begin{center}
\mbox{\epsfig{file=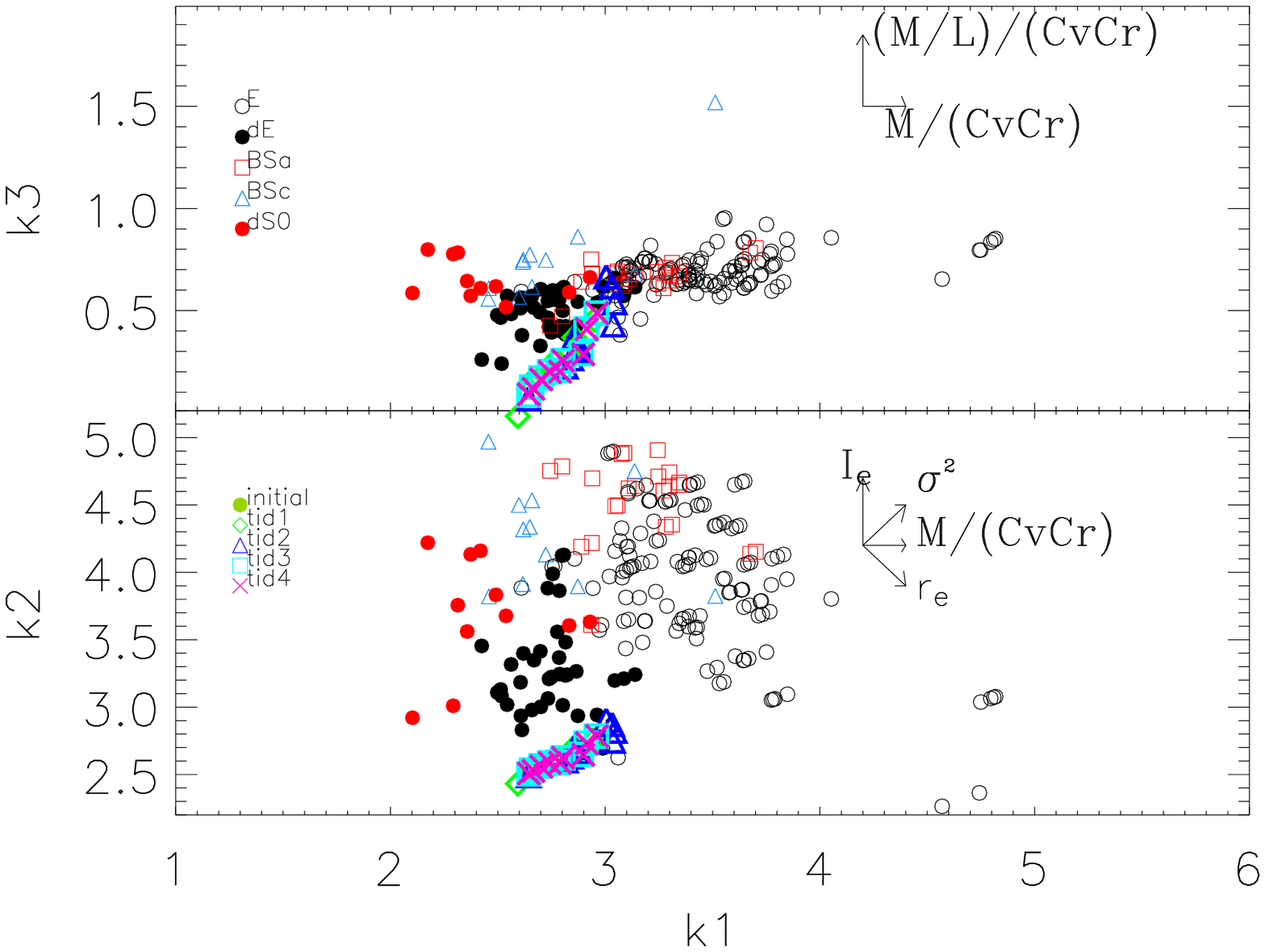,width=8cm}}
\mbox{\epsfig{file=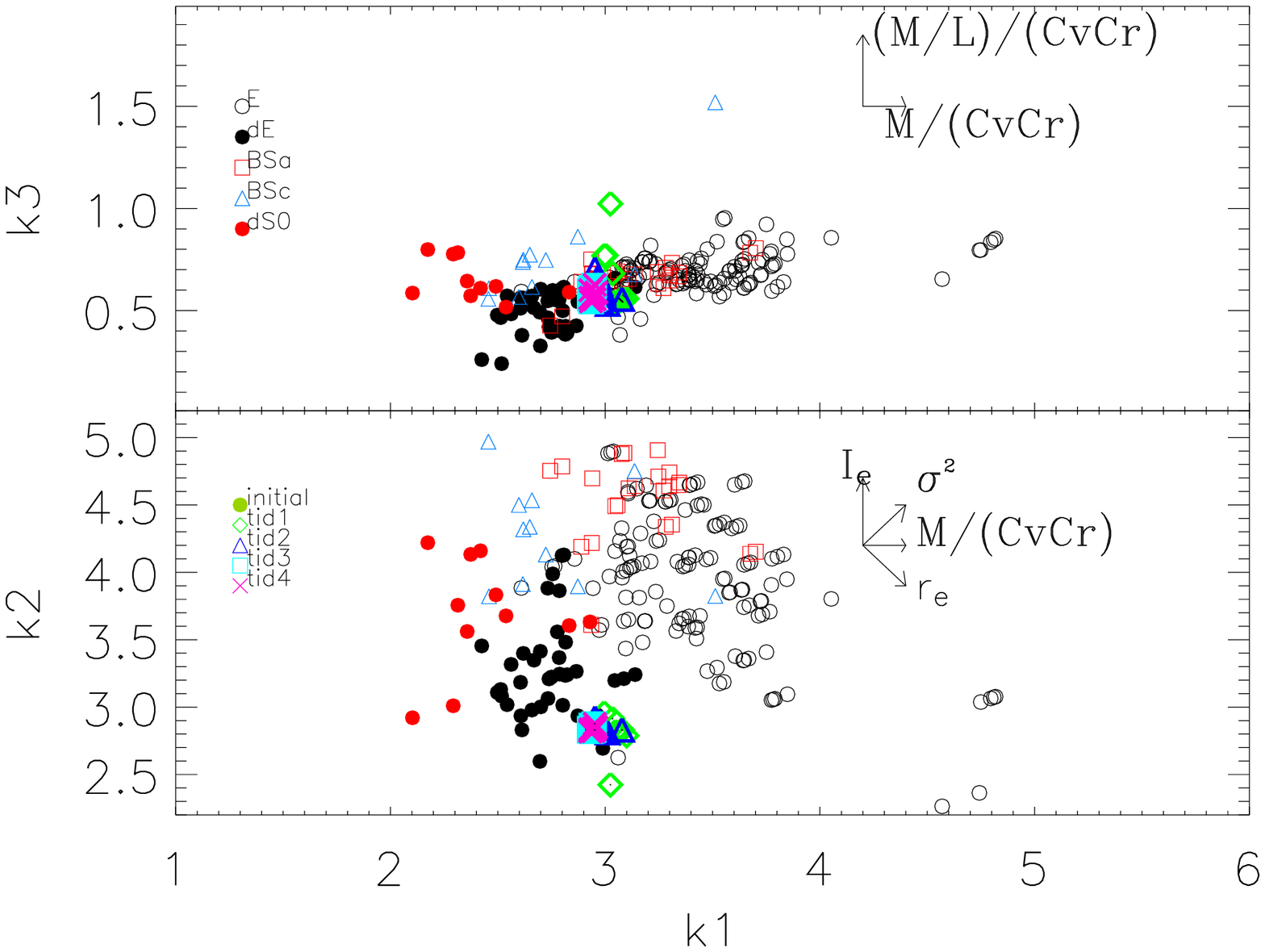,width=8cm}}
\end{center}
\caption{Results for the models TID5 and TID6 presented in the $\kappa$-space. Symbols and colours are as in Figure\ref{Fig:fig2}. }
\label{Fig:fig30}
\end{figure}

\begin{table}
\begin{center}
\caption{Maximun rotation velocity and velocity dispersion of the initial and final models and mean $V/\sigma$ inside the effective radius for the final model.\label{tab3}}
\begin{tabular}{@{}cccccc}
\hline
{\bf $Model$}& {\bf $V_{max, initial}$} & {\bf $V_{max, final}$} & {\bf $\sigma_{max, initial}$} & {\bf $\sigma_{max,final}$}& {\bf $(V/\sigma)_{final}$} \\
             & (km s$^{-1}$) & (km s$^{-1}$) & (km s$^{-1}$) & (km s$^{-1}$) \\
\hline
TID1&  181.1 & 70.3 & 46.6 & 49.6 & 0.034\\
TID2&  252.6 & 94.8 & 93.7 & 92.5 & 0.020\\
TID3&  284.0 & 136.3& 194.3 & 163.9& 0.013\\
TID4&  284.0 & 138.7 & 192.8 & 170.6 & 0.015\\
TID5&  149.3 & 111.2 & 37.0 & 46.9 & 0.017\\
TID6&  149.3 & 57.9 & 37.0 & 68.9 & 0.021\\
TID7&  149.3 & 21.4 & 37.0 & 78.1 & 0.014\\
TID8&  149.3 & 31.6 & 37.0 & 72.6 & 0.019\\
GST1&  247.6 & 89.1 & 89.0 & 90.8& 0.032\\
GST2&  179.9 & 87.4 & 50.3 & 50.6& 0.018\\
\hline  
\end{tabular}
\end{center}
\end{table}

\subsubsection{Kinematical features of the remnants}
Figure~\ref{Fig:fig20} shows the rotation curve and the velocity dispersion profile of the initial and final remnants of all simulations. Table 3 gives the maximum rotation and velocity dispersion of the initial and final models. Notice that the rotation velocity of all models are significantly smaller in the external regions at the end of the simulations than in the initial conditions. In contrast, the external region of the galaxies show also in all simulations larger velocity dispersion than at the beginning. This is caused because tidal interactions are an efficient mechanism for heating the thin disc of spiral galaxies (see e.g. \cite{aguerri01, eliche06}). The central velocity dispersion at the end of the simulations depends on the models. Those models with an initial small bulge show very similar central velocity dispersion at the beginning and at the end of the simulations (TID1, TID2). Models with large bulge at the initial conditions (TID3 and TID4) show smaller central velocity dispersion at the end. This is due to the final bulges of this models suffered an expansion. On the other hand, the models without initial bulge (TID5-TID8) show larger central velocity dispersion at the end, again due to the tidal heating of the disc and the absence of a stabilising central concentration like a bulge. Retrograde models seem to present slightly higher values of the central velocity dispersion. Table 3 also shows the $V/\sigma$ ratio measured within an aperture equal to $r_{e}$. Notice that the final remnants are in all cases dominated by random motions. Similar kinematics features have also been seen in other simulations of fast galaxy encounters in the literature (see Kronberger et al. 2006).

The kinematical characteristics of the models with simulated tidal stripping (GST1 and GST2) do not differ largely from those similar simulations without tidal stripping (TID1 and TID2). This is because the gas-particles only account for a few percent (10$\%$ in our simulations) of the total mass of the disc in the galaxies.

\subsection{The Fundamental Plane of the tidal remnants}

We have analysed the location of the simulated galaxies in the FP after each tidal encounter. We have measured the required quantities $\mu_{e}$, $r_{e}$ and $\sigma_{o}$, in the simulations. 

\subsubsection{Decomposition of the surface mass density profiles}

Similar to real galaxies, we have performed the decomposition of the surface mass density profiles in the different galactic components of the initial modelled galaxy and the remnants after each interaction. This decomposition was done using one or two components depending if the systems are dominated by rotation or pressure in their most external regions. Thus, systems with $v/\sigma < 1$ at all radial distances were considered as hot systems. In contrast, those systems with $v/\sigma>1$ at large radial distances were considered as rotation supported systems in their outer most regions. Thus, the surface mass density profiles of the galaxies with a rotating component in their outermost region were fitted with two components: bulge and disc. The bulge component dominates the surface mass density profiles in the innermost regions of the models, and was represented by a S\`ersic law (\cite{sersic68}). This component represents the hottest part of the galaxy being dominated by random motions, and it will be called ``bulge'' similar to real galaxies. The disc component dominates the outermost regions of the mass density profiles of the galaxies, and was modelled by a Freeman exponential law (Freeman et al. 1970). The surface mass density profiles of those models dominated by random motions at all radial distances were fitted by one-component model described by a S\`ersic law. 

The free parameters of the models were fitted using a Levenberg-Marquard nonlinear fitting routine and were determined by minimising the $\chi^{2}$ of the fit similar to real galaxies (Trujillo et al. 2001, 2002; Aguerri \& Trujillo 2002; Aguerri et al. 2004, 2005a).

The decompositions of the surface mass density profiles provided two of the parameters of the hot component of the galaxy  needed for the FP, the effective radius (r$_{e}$) and the effective surface mass density ($\Sigma_{e}$). The third parameter (central velocity dispersion) was measured from the velocity dispersion profiles of the galaxies in an aperture of radius equal to 0.2$\times r_{e}$ of the hot galaxy component. All these FP parameters were obtained for 9 different points of view in each system.

As pointed out before, the discs of the galaxies are strongly truncated. This is reflected in the scale-length of the discs which is reduced in all cases up to 40-50$\%$. In contrast, the variation in the effective radius of the bulges depends on the prominence of this structure in the initial model. Thus, the scale of bulges in models with small $B/D$ ratio (TID1 and GST1) was very small. In contrast, galaxies with large $B/D$ ratio (TID3 and TID4) have larger scales of the bulge at the end of the interactions. This behaviours of the bulge and disc structural parameters produce that the B/D ratio of the tidally truncated galaxies is larger than the initial one.

\subsubsection{Location of tidal remnants in the FP}

To locate the modelled galaxies in the observed FP we need to scale them using an appropriate mass-to-light ratio. This $M/L$ was obtained by imposing that the hot systems of the simulated initial galaxy and remnants lied on the edge-on view of the FP described by bright E galaxies. With this $M/L$ we have compared the location of those systems in the face-on view of the FP. We have varied the imposed $M/L$  according with the rms showed in the edge-on view of the FP of early-type bulges (TID2, TID3, TID4 and GST2) and late-type bulges (TID1 and GST1). The result of this variation in the face on view of the FP is given as an error bar in the following FP figures. Figure \ref{Fig:fig9} shows the location in the $\kappa$-space of the observed galaxies in the Coma cluster presented in \S~\ref{sec:FP}, together with the hot components corresponding to the TID1, TID2 and TID3 models. We find that those simulations with initial conditions similar to an early-type spiral (TID3 model) are located close to the face-on view of the FP described by bulges of early-type and E galaxies. In contrast, the remnants of the models with small B/D ratio at the beginning (TID1 model) are located in a different position in the face-on view of the FP defined by E galaxies, similar to dE ones. The remnants from TID2 are between TID1 and TID3 models. 

Models GST1 and GST2 have a gas-stripping event at the beginning of the simulations. This event is supposed to erase all gas from the galaxy in a rather short amount of time. This modifies the profile of our initial conditions producing initial bulges with larger scales than in similar models without gas-stripping (TID1 and TID2 models). The final remnants for these models are located at different positions in the face-on view of the FP than those from TID1 and TID2 simulations (see Fig.~\ref{Fig:fig10}). In fact, we find a marginally better agreement with the locus of dEs in the model GST1. 

All these models with galaxies with bulge at the beginning can explain the location in the face-on view of the FP of dwarf galaxies with central velocity dispersion larger than $\approx 50$ km s$^{-1}$. But, dwarf galaxies with smaller central velocity dispersions are yet not well reproduced.

The $M/L$ ratio for the four models without initial bulge component (TID5-TID8) were obtained in a different way. In this case the initial models represent pure disc galaxies, and it has no sense to locate these discs in the FP. We have obtained the $M/L$ ratio by locating the initial discs of these models in the Tully-Fisher (Tully \& Fisher (1977); TF) relation followed by spiral galaxies. We have the maximum rotational velocity for these discs, and following the Tully-Fisher calibration of Sakai et al. (2000) in the I band we have obtained the $M/L$. 

Figure \ref{Fig:fig30} shows the location of the remnants from TID5 and TID6 models in the edge-on (upper panel) and face-on (lower panel) for the FP. The final remnants from these models show smaller central velocity dispersions than the other simulations, and are located in the region occupied by dE with the smallest central velocity dispersions. These results do not depend on the orbit of the interactions. Thus, the models TID7 and TID8, following retrograde orbits, show similar results.

\section{Discussion}

\begin{figure}
\begin{center}
\mbox{\epsfig{file=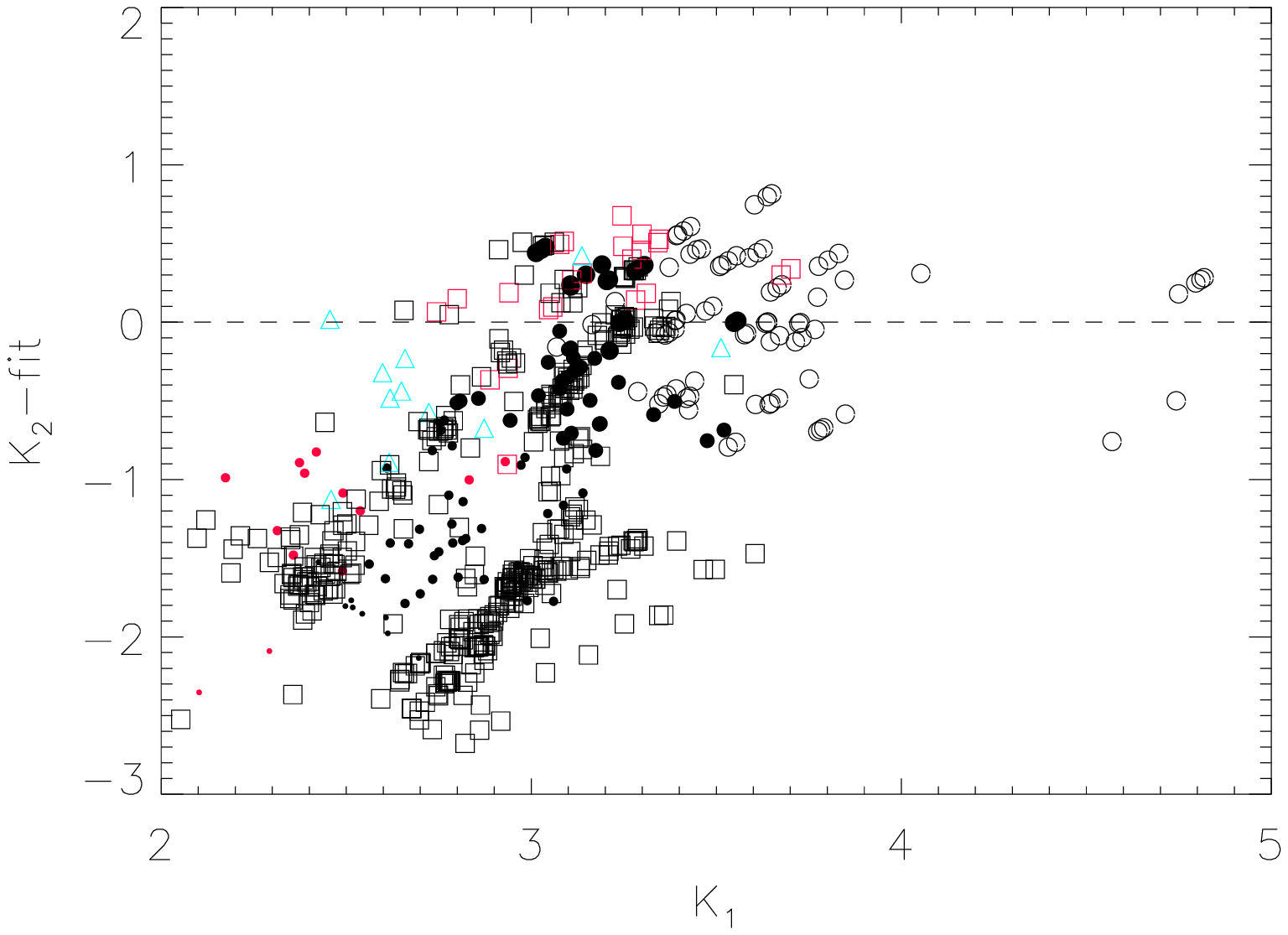,width=9cm}}
\end{center}
\caption{Residuals of the face-on FP in the $\kappa$ representation of the different types of galaxies in the Coma cluster: bright E(empty circles), dE (black circles), early-type bulges (red squares), late-type bulges (blue triangles), and dS0 bulges (red circles). The empty squares represent the simulated galaxies from models TID1-TID8, GST1 and GST2. }
\label{Fig:fig11}
\end{figure}

The simulations presented in the present paper indicate that fast tidal interactions are very efficient transforming spiral disc galaxies into dwarf like systems. This mechanism produces a large mass loss, specially from the outer parts of the galaxies, and locate the tidal remnants out from the face-on view of the FP described by bright E galaxies. Figure \ref{Fig:fig11} shows the residuals of the observed galaxies and the remnants of the present models respect to the face-on view of the FP described by E bright galaxies. Notice the tidal remnants from the present simulations are located in the same position of the FP than dE galaxies. Galaxies with large B/D-ratios at the beginning are located after four tidal fast interaction near to the loci occupied by bulges of early-type galaxies. In contrast, galaxies with small B/D-ratio or without initial bulge are located at the end of the simulation at the position of the dE galaxies with small central velocity dispersion. This indicates that dE galaxies can be formed by the tidal disruption of late-type disc galaxies. In contrast, the tidal disruption of early-type spirals would produce a final galaxy similar to elliptical or S0 galaxies.

This is not the unique mechanism that can explain the location of dwarf galaxies in the face-on view of the fundamental plane. Semi-analytical models of dwarf galaxy evolution by supernova-driven-gas-loss can also explain the position of dwarf galaxies in the face-on view of the FP (see de Rijcke et al. 2005). This could indicate that the origin of the dwarf population could be diverse. Thus, several past studies have shown that dwarf galaxies are not an homogeneous population of objects. Recently, Lisker et al. (2007) have analysed a large sample of dwarf elliptical galaxies in the Virgo cluster, concluding that they can be grouped in three main families with different dynamical and  morphological properties. They found that non-nucleated dE show shapes like thick discs and are not centrally clustered. This indicates that they constitute a non-relaxed galaxy population, and could be formed by tidal interactions as our simulations show. But, how can we know the progenitors of dwarf galaxies?, are there any observational measurement for knowing the tidal origin of dwarf galaxies?.

\subsection{Bar parameters in dwarf galaxies}

\begin{figure}
\begin{center}
\mbox{\epsfig{file=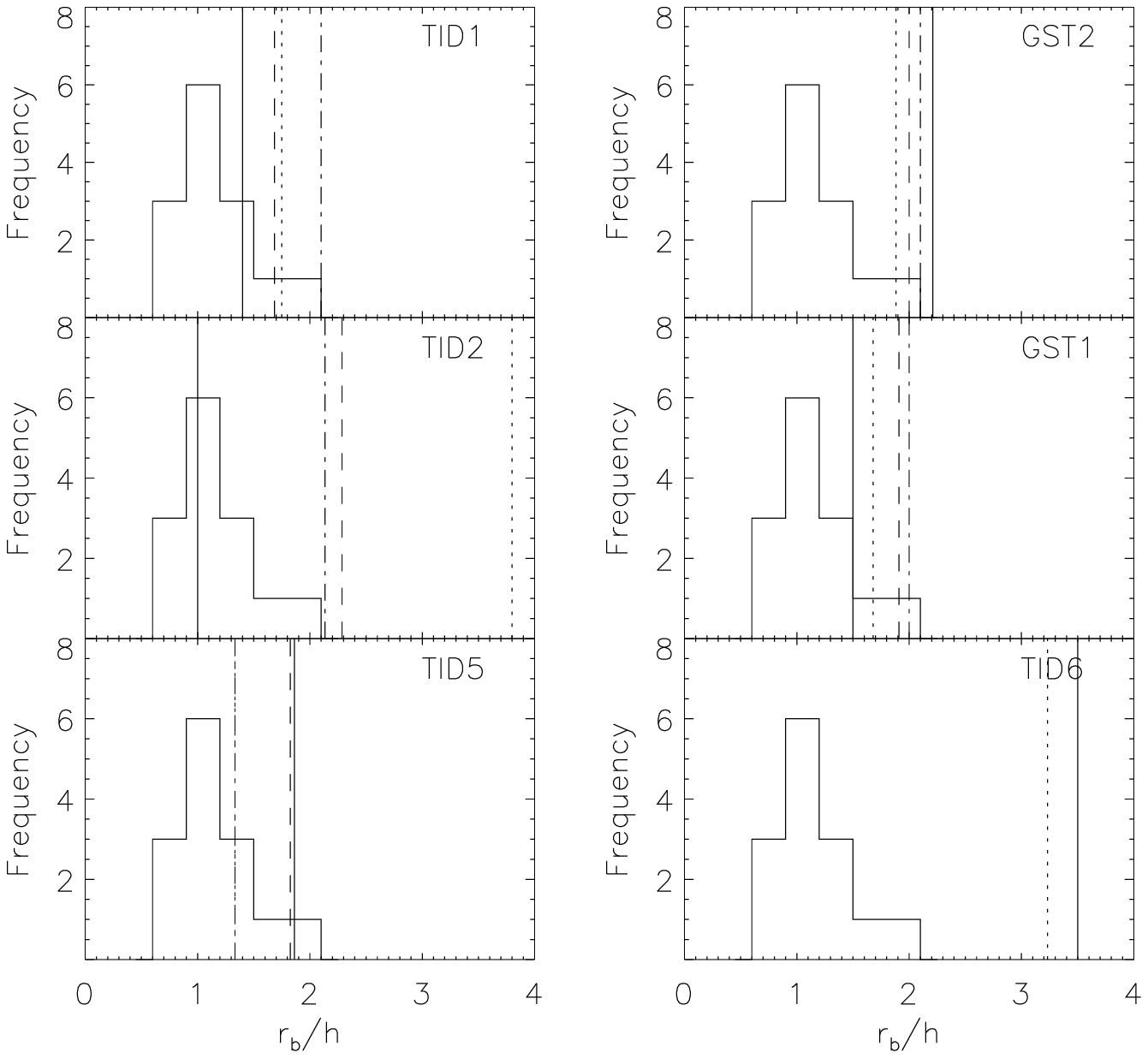,width=9cm}}
\end{center}
\caption{Ratio between the length of the bars ($r_{b}$) and the scale length of the discs ($h$) in observed bright unperturbed barred galaxies (histogram) and simulations with bars formed by tidal interactions (vertical lines). The different lines represent bars formed after 1 (full line), 2 (dotted line), 3 (dashed line) or 4 (dashed-dot-dot line) tidal encounters.}
\label{Fig:fig8}
\end{figure}

As we mentioned in the previous sections, one of the most prominent features that tidal interactions induce in the discs of the galaxies is the presence of stable and prominent bars. Those bar features appeard only in those models with small bulge component and in interactions following prograde orbits. Retrograde orbits inhibits bar formation even in galaxies with small bulge. We can analyse the properties of these bars and compare them with similar properties of bars observed in bright galaxies without signs of interaction. Numerical simulations of bar formation find that the mean ratio between the bar length and the scale length of the galaxy discs ($r_{b}/h$) is well constrained, being 1.0-1.2 (O'Neill \& Dubinski 2003; Valenzuela \& Klypin 2003). This is in agreement with observational measurements of this quantity in bright unperturbed barred galaxies ($<r_{b}/h> = 1.21 \pm 0.08$; Aguerri et al. 2005c). We have measured this ratio in the bars induced in our models by tidal interactions. Figure \ref{Fig:fig8} shows the $r_{b}/h$ distribution on a sample of 14 bright barred galaxies (Aguerri et al. 2005c) and the values of this ratio for the bars formed in our simulations. It is clear than the tidally induced bars have larger $r_{b}/h$ values than observed bars. This could be an observational parameter that could be measured in dwarf barred galaxies in order to distinguish between tidally induced bars and those formed by internal discs instabilities.

There are few measurements of the ratio $r_{b}/h$ in the literature for dwarf galaxies. Nevertheless, NGC4431 is one exception. This is a barred dwarf galaxy located in the Virgo cluster and analysed by Corsini et al. (2007). The surface brightness profile of this galaxy was decomposed in three components (bulge, disc and bar) resulting $r_{b}/h=1.4$, similar to the values obtained in our simulations. Therefore, the bar of this galaxy could be formed in a past tidal interaction.

\subsection{Structural parameters of tidal dwarf galaxies}

\begin{figure}
\begin{center}
\mbox{\epsfig{file=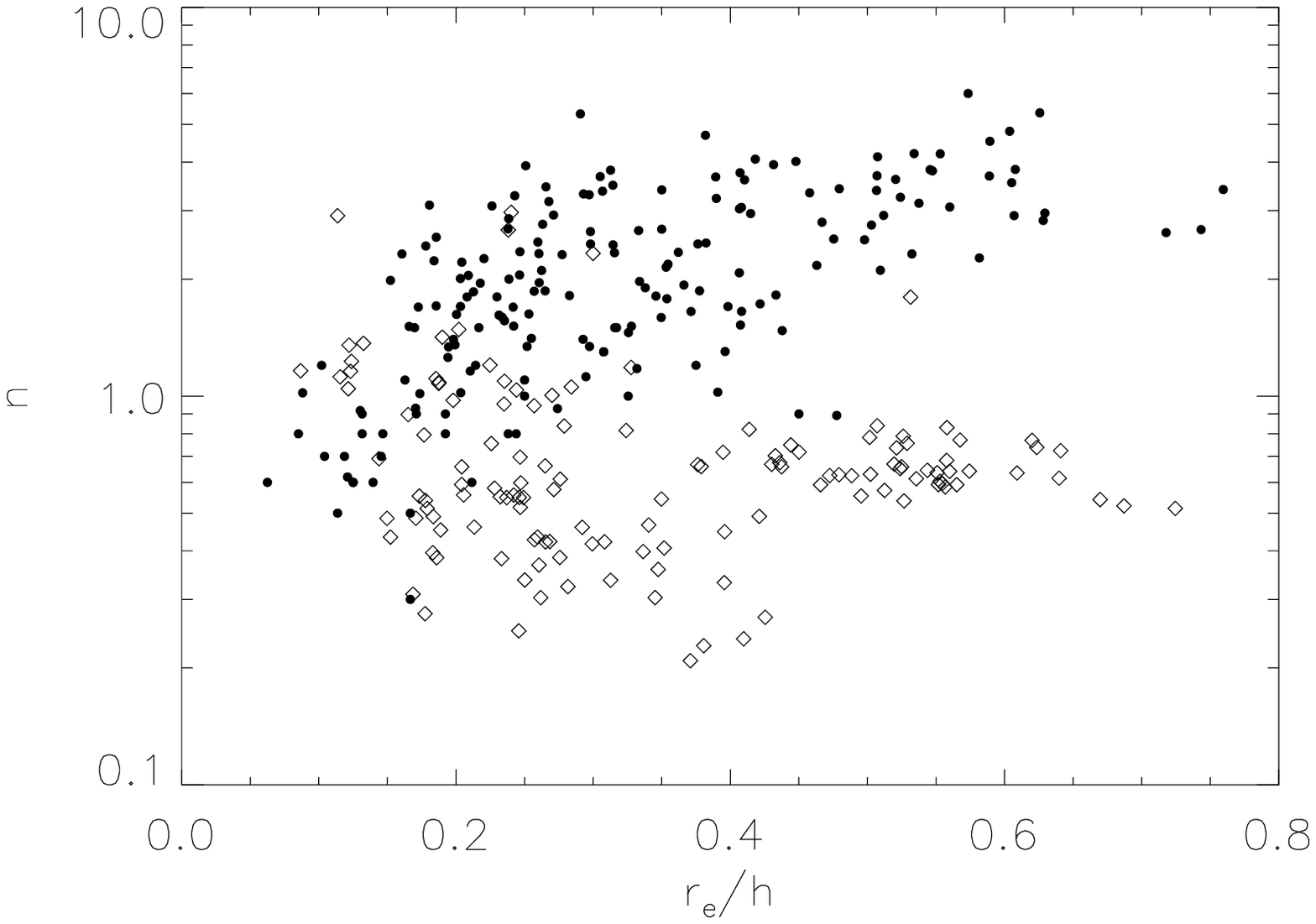,width=9cm}}
\end{center}
\caption{$r_{e}/h$ vs h for observed bright undisturbed galaxies (black points) and simulated dwarf galaxies formed by harassment of bright disc ones (diamonds). }
\label{Fig:fig12}
\end{figure}

The analysis of the structural parameters of the dwarf galaxies can also give clues about their tidal origin. Photometric decomposition of bright and unperturbed galaxies have shown several correlations between bulge and disc structural parameters (e.g. \cite{dejong96}; \cite{khosroshahi00}; \cite{macarthur03}; \cite{mollenhoff04}; \cite{hunt04}; \cite{mendezabreu08}) . In particular, the ratio between the disc and bulge scales ($r_{e}/h$) is correlated with the S\`ersic $n$ parameter in real disc galaxies (e.g. MacArthur et al. 2003; M\'endez-Abreu et al. 2008). Thus, those bulges with large $n$ values also show large $r_{e}/h$ ratios. We have compared the correlation observed in real galaxies with our simulations. Figure \ref{Fig:fig12} shows the correlation between $r_{e}/h$ and $n$ for a large sample of bright disc galaxies from the surveys of MacArthur et al. (2003) and M\'endez-Abreu et al. (2008). We have also overploted in this figure the same quantities for our simulated dwarf-like objects. Notice that dwarf galaxies formed by fast tidal interactions are most of them outside the trend described by real galaxies. Thus, for a fixed value of $n$ tidally formed dwarf galaxies have, in general, larger $r_{e}/h$ ratios than bright disc spirals. This is because, as we discuss in the previous section, the scales of the discs were strongly truncated due to tidal encounters. The search of outliers in the $r_{e}/h$ vs $n$ relation could provide dwarf systems formed by transformed bright disc galaxies by fast tidal encounters. 

We have the bulge and disc structural parameters of the dwarf barred galaxy NGC4431. This galaxy shows a S\`ersic index $n=1.5$ and $r_{e}/h=0.44$ (Corsini et al. 2007). According with Fig. \ref{Fig:fig12} this galaxy would be in the limit of the region described by the observed bright disc galaxies. This would be another indication that NGC4431 could be a dwarf galaxy formed by stripped bright disc galaxies.

\subsection{Metallicity relations}

Early-type galaxies follow the well known luminosity-metallicity and velocity dispersion-metallicity relations (see e.g. Tremonti et al. 2004 and references therein). Thus, more metallic galaxies are those with larger central velocity dispersions and more luminous ones. Tidal stripping interactions should not change the metallicity of the galaxies. Indeed, we have also seen that the central velocity dispersion of tidal remnants are not very different from the initial conditions. In contrast, due to the large mass lost, tidally formed dwarf galaxies have lower luminosities than their progenitors. This would imply that low-mass systems formed by harassment of bright disc galaxies could be out of the luminosity-metallicity relation displayed by bright galaxies. In contrast, those dwarf galaxies could lie inside the relation between the metallicity and their central velocity dispersion.

 \subsection{Group and Cluster fast tidal interactions}

We have made some attempts to apply the tidal impulse approximation to model encounters with group and cluster potentials. Mastropietro et al. (2005) simulate the harassment of disc galaxies falling onto a cluster environment by re-simulating a cluster formed in a cosmological simulation. We have tried to model the harassment of one of our bulgeless galaxies through the tidal impulse approximation. However, two important problems arise: The most typical orbital configuration from the cosmological simulations (Tormen 1997) are such that the injection of energy is too low to produce significant effects in the discs. This is so because the typical orbits imply wide impact parameters on non-parabolic orbits. The second problem is that under this circumstances no longer applies the impulse approximation (i.e., the interaction is not fast compared with the dynamic time of the galaxy), and the approach followed in this paper is no longer applicable.

Nevertheless we have modelled one encounter with  higher mass ratio (1:10). This model (called TID10) simulates the interaction of a bulgeless galaxy with the potential of a group. The orbital parameters chosen for this encounter were: $b = 6$, $V = 4.6$ and $R_{peri}= 1$. This orbit is not a typical orbit (Tormen 1997) but the parameters are such that the increment of energy to each particle is comparable to the ones in galaxy-galaxy interactions. In fact, the results are quite similar to the simulations TID5-TID8 (see figure 6 for the mass change). The final half mass radii of the halo and disc components were $ r_{e,halo}=0.532$ and $ r_{e,disc}=0.370$, respectively. The final  maximum rotational velocity and central velocity dispersion were $ 37$ km s$^{-1}$, and $ 55.6$ km s$^{-1}$, respectively. All these values are similar to the ones shown in Tab. 2 for the models without bulge in the initial conditions (TID5-TID8). Indeed, the final position of these remnants in the FP was similar to similar models (TID5-TID8) showed in the present paper. However, we must stress that the orbital configuration is not typical at all, compared with those obtained from cosmological simulations (Tormen 1997).

We would like also to point that the models showed in this subsection are very simple. The interaction of a galaxy with the global cluster or group potentials should take into account NFW-type tidal fields from the potential. This kind of interactions are out of the main scope of the present paper where galaxy-galxy interactions were shown. We are planning new cosmological simulations considering both the galaxy-galaxy and galaxy-cluster potential interactions.

\section{Conclusions}

In the present paper we have shown the FP of 215 hot systems (ellipcicals and bulges) located in the Coma clusters. The edge-on view of the FP shows that bright E galaxies and early-type bulges are located in the same plane showing similar scatter. In contrast, dwarf galaxies and late-type bulges have larger scatter. The differences are more pronounced in the face-on view of the FP. Thus, dwarf galaxies and bulges of late-type galaxies are located perpendicular to the relationship defined by E and early-type bulges. 

We have run high-resolution numerical simulations in order to explain the different location of dwarf galaxies in the face-on view of the FP. We have tested if fast galaxy-galaxy tidal interactions can transform bright spiral galaxies in early-type dwarf systems and can explain their different location in the FP. The simulations have  disc-like galaxies with different bulge-to-disc mass rations as initial conditions. Thus, these initial disc galaxies represent the full family of Hubble type galaxies, from early- to late-type ones.  

Tidal interactions turned out to be very efficient for removing stars from the galaxies, specially from the outermost regions (disc and dark matter halo) where the stars are less bounded. In contrast, particles from bulges were not removed. After several tidal interactions the galaxies have lost 50-80$\%$ of their total mass and 30-60$\%$ of the luminous matter. The discs and haloes turned to be strongly truncated. In particular, the structure of the dics were largely affected. Their scale parameters were 40-50$\%$ shorter than the initial ones. Those models with the smallest B/D ratio retain a short and thick disc. In contrast, the model with the largest B/D ratio have no disc at the end, being a roundish dwarf system. The tidal interactions produce bar-like structures in the galaxy which are stable until the end of the models. The bars were more prominent in those models with smaller bulges at the beginning of the simulations. The formation of bar-like structures depends on the orbital configuration of the encounters. Thus, bars are easily formed in prograde interactions. In contrast, retrograde orbits seem to inhibit the formation of bars even in those simulations without initial bulge component. Similar results have been obtained for two models in which we have simulated a gas stripping interaction before tidal encounters.

We have analysed the location of the dwarf-like galaxies remnants in the face-on view of the FP. Their final positions depend on the initial conditions of the simulations. Thus, those dwarf galaxies with progenitors with large B/D ratio are closer to the location of early-type bulges and Es in the face-on view of the FP. In contrast, the remnants of late-type galaxies are located in the region occupied by dE in the face-on view of the FP. The final location of the remmants in the FP does not depend on the orbit orientation of the encounters.

We have analysed several observational parameters in order to identify these tidal dwarf galaxies. The measurement of the  $r_{b}/h$ ratio, the relation between $r_{e}/h$ and $n$, and the mass-metallicity and velocity dispersion-metallicity relations could provide information about the possible tidal origin of observed dwarf galaxies. Therefore, tidal dwarf systems would have larger $r_{b}/h$ ratio than the observed one in bright barred galaxies. They would also be outliers in the $r_{e}/h$-$n$ and luminosity-metallicity relations. In contrast, they would follow the velocity dispersion-metallicity relation.

\begin{acknowledgements}
We would like to thank useful comments from L. Mayer and B. Moore during the writing process of the present manuscript. We would also like to thank to the anonymous referee for the usefull comments given. We wish to thank the BSC in Barcelona for computing facilities at the Mare Nostrum supercomputer and at the La Palma node. ACGG acknowledges financial support  by the regional government of Madrid through the ASTROCAM Astrophysics network (S­0505/ESP­0237). JALA also acknowledges financial support by the project AYA2007-67965-C03-01.
 
\end{acknowledgements}


\begin{thebibliography}{}

\bibitem[Aceves 
\& Vel{\'a}zquez 2005]{aceves05} Aceves, H., \&
  Vel{\'a}zquez, H.\ 2005, \mnras, 360, L50  

\bibitem[Aguerri et al. 2005a]{aguerri05a} Aguerri, J.~A.~L., 
Iglesias-P{\'a}ramo, J., V{\'{\i}}lchez, J.~M., Mu{\~n}oz-Tu{\~n}{\'o}n, 
C., \& S{\'a}nchez-Janssen, R.\ 2005a, \aj, 130, 475 

\bibitem[Aguerri et al. 2005b]{aguerri05b} Aguerri, J.~A.~L., 
Gerhard, O.~E., Arnaboldi, M., Napolitano, N.~R., Castro-Rodriguez, N., \& 
Freeman, K.~C.\ 2005b, \aj, 129, 2585 

\bibitem[Aguerri et al. 2005c]{aguerri05c} Aguerri, J.~A.~L., Elias-Rosa, N., Corsini, E.~M., \& Mu{\~n}oz-Tu{\~n}{\'o}n, C.\ 2005c, \aap, 434, 109 

\bibitem[Aguerri et al. 2004]{aguerri04} Aguerri, J.~A.~L., 
Iglesias-Paramo, J., Vilchez, J.~M., \& Mu{\~n}oz-Tu{\~n}{\'o}n, C.\ 2004, 
\aj, 127, 1344 



\bibitem[Aguerri \& Trujillo 2002]{aguerri02} Aguerri, J.~A.~L., \& Trujillo, I.\ 2002, \mnras, 333, 633 


\bibitem[Aguerri et al. 2001]{aguerri01} Aguerri, J.~A.~L., Balcells, M., \& Peletier, R.~F.\ 2001, \aap, 367, 428 

\bibitem[Arnaboldi et al. 2002]{arnaboldi02} Arnaboldi, M., et 
al.\ 2002, \aj, 123, 760 

\bibitem[Athanassoula et al. 2005]{athanassoula05} Athanassoula, E., 
Lambert, J.~C., \& Dehnen, W.\ 2005, \mnras, 363, 496 

\bibitem[Bekki 1998]{bekki98} Bekki, K.\ 1998, \apj, 496, 713 

\bibitem[Bender et al. 1992]{bender92} Bender, R., Burstein, 
D., \& Faber, S.~M.\ 1992, \apj, 399, 462 

\bibitem[Bernardi et al. 2003]{bernardi03} Bernardi, M., et al.\ 
2003, \aj, 125, 1866 

\bibitem[Binggeli \& Cameron 1991]{binggeli91} Binggeli, B., \& 
Cameron, L.~M.\ 1991, \aap, 252, 27 

\bibitem[Binney \& Tremaine 1987]{binneyT87} Binney, J., \& Tremaine, S.~1987, Galactic Dynamics, Princeton University Press, Princeton

\bibitem[Boselli et al. 2008]{boselli08} Boselli, A., Boissier, 
S., Cortese, L., \& Gavazzi, G.\ 2008, \apj, 674, 742

\bibitem[Boylan-Kolchin, Ma, Quataert]{boylan06} Boylan-Kolchin, M., Ma, C.~P., Quataert, E.\ 2006,  \mnras, 369, 1081 

\bibitem[Caldwell \& Bothun 1987]{caldwell87} Caldwell, N., \& 
Bothun, G.~D.\ 1987, \aj, 94, 1126 

\bibitem[Capelato et al. 1995]{capelato95} Capelato, H.~V., de 
Carvalho, R.~R., \& Carlberg, R.~G.\ 1995, \apj, 451, 525 

\bibitem[Conselice et al. 2001]{conselice01} Conselice, C.~J., 
Gallagher, J.~S., III, \& Wyse, R.~F.~G.\ 2001, \apj, 559, 791 

\bibitem[Corsini et al. 2007]{corsini07} Corsini, E.~M., 
Aguerri, J.~A.~L., Debattista, V.~P., Pizzella, A., Barazza, F.~D., 
\& Jerjen, H.\ 2007, \apjl, 659, L121 


\bibitem[Davies et al. 2008]{davies08} Davies, K. I., Roberts, S. \& Sabatini, s. \ 2008, \mnras, in press

\bibitem[Davies et al. 1983]{davies83} Davies, R.~L., 
Efstathiou, G., Fall, S.~M., Illingworth, G., \& Schechter, P.~L.\ 1983, 
\apj, 266, 41 





\bibitem[Dekel \& Silk 1986]{dekel86} Dekel, A., \& Silk, J.\ 
1986, \apj, 303, 39 


\bibitem[de Jong 1996]{dejong96} de Jong, R.~S.\ 1996, \aap, 313, 45 


\bibitem[de Rijcke et al. 2005]{derijcke05} de Rijcke, S., 
Michielsen, D., Dejonghe, H., Zeilinger, W.~W., \& Hau, G.~K.~T.\ 2005, 
\aap, 438, 491 

\bibitem[Djorgovski \& Davis 1987]{djorgovski87} Djorgovski, S., \& Davis, M.\ 1987, \apj, 313, 59 

\bibitem[D'Onofrio et al. 2006]{donofrio06} D'Onofrio, M., 
Valentinuzzi, T., Secco, L., Caimmi, R., 
\& Bindoni, D.\ 2006, New Astronomy Review, 50, 447 

\bibitem[Dressler 1987]{dressler87} Dressler, A.\ 1987, \apj, 
317, 1 

\bibitem[Eliche-Moral et al. 2006]{eliche06} Eliche-Moral, M. C., Balcells, M., Aguerri, J. A. L., \& Gonz\'alez-Garc\'{\i}a, C., 2006, \aap, 457, 91.

\bibitem[Faber \& Jackson 1976]{faber76} Faber, S.~M., \& Jackson, R.~E.\ 1976, \apj, 204, 668 

\bibitem[Falc{\'o}n-Barroso et al. 2002]{falconbarroso02} Falc{\'o}n-Barroso, J., Peletier, R.~F., \& Balcells, M.\ 2002, \mnras, 335, 741 

\bibitem[Feldmeier et al. 2004]{feldmeier04} Feldmeier, J.~J., 
Ciardullo, R., Jacoby, G.~H., \& Durrell, P.~R.\ 2004, \apj, 615, 196 

\bibitem[Forbes \& Ponman 1999]{forbes99} Forbes, D.~A., \& Ponman, T.~J.\ 1999, \mnras, 309, 623

\bibitem[Freeman 1970]{freeman70} Freeman, K.~C.\ 1970, \apj, 
160, 811 

\bibitem[Geha et al. 2003]{geha03} Geha, M., Guhathakurta, 
P., \& van der Marel, R.~P.\ 2003, \aj, 126, 1794 

\bibitem[Geha et al. 2002]{geha02} Geha, M., Guhathakurta, 
P., \& van der Marel, R.~P.\ 2002, \aj, 124, 3073 

\bibitem[Gerhard et al. 2007]{gerhard07} Gerhard, O., Arnaboldi, M., Freeman, K.~C., Okamura, S., Kashikawa, N., \& Yasuda, N.\ 2007, \aap, 468, 815 

\bibitem[Gerhard et al. 2005]{gerhard05} Gerhard, O., Arnaboldi, 
M., Freeman, K.~C., Kashikawa, N., Okamura, S., 
\& Yasuda, N.\ 2005, \apjl, 621, L93 

\bibitem[Gnedin et al.(1999)]{Gnedin99} Gnedin, O.~Y., 
Hernquist, L.~\& Ostriker, J.~H.\ 1999, \apj, 514, 109 

\bibitem[Gonz\'alez-Garc\'\i a, Aguerri, Balcells, 2005]{GGAB05} Gonz\'alez-Garc\'\i a, A.~C., Aguerri, J.~A.~L.~\& Balcells, M.\ 2005, \aa, 444, 803

\bibitem[Gonz\'alez-Garc\'\i a \& van Albada]{GGvA03} Gonz\'alez-Garc\'\i a, A.~C.~\& van Albada, T.~S.\ 2003, \mnras, 342, L36

\bibitem[Graham \& Guzm{\'a}n 2003]{graham03} Graham, A.~W., \& 
Guzm{\'a}n, R.\ 2003, \aj, 125, 2936 

\bibitem[Grebel et al. 2003]{grebel03} Grebel, E.~K., 
Gallagher, J.~S., III, \& Harbeck, D.\ 2003, \aj, 125, 1926 


\bibitem[Guti{\'e}rrez et al. 2004]{gutierrez04} Guti{\'e}rrez, 
C.~M., Trujillo, I., Aguerri, J.~A.~L., Graham, A.~W., \& Caon, N.\ 2004, 
\apj, 602, 664 

\bibitem[Guzman et al. 1993]{guzman93} Guzman, R., Lucey, 
J.~R., \& Bower, R.~G.\ 1993, \mnras, 265, 731 

\bibitem[Held et al. 1992]{held92} Held, E.~V., de Zeeuw, T., 
Mould, J., \& Picard, A.\ 1992, \aj, 103, 851 

\bibitem[Hunt et al. 2004]{hunt04} Hunt, L.~K., Pierini, D., \& Giovanardi, C.\ 2004, \aap, 414, 905 

\bibitem[Jorgensen et al. 1996]{jorgensen96} Jorgensen, I., Franx, 
M., \& Kjaergaard, P.\ 1996, \mnras, 280, 167

\bibitem[Kauffmann \& White 1993]{kauffmann93} Kauffmann, G., \& White, S.~D.~M.\ 1993, \mnras, 261, 921 

\bibitem[Khosroshahi et al. 2000]{khosroshahi00} Khosroshahi, H.~G., 
Wadadekar, Y., \& Kembhavi, A.\ 2000, \apj, 533, 162 

\bibitem{} King, I.~R.,~1966, AJ, 71, 64

\bibitem[Kormendy 1985]{kormendy85} Kormendy, J.\ 1985, \apj, 
295, 73 

\bibitem[Kormendy 1977]{kormendy77} Kormendy, J.\ 1977, \apj, 
218, 333 

\bibitem[Kronberger et al.(2006)]{kronberger06} Kronberger, T., Kapferer, W., Schindler, S., B{\"o}hm, A., Kutdemir, E., \& Ziegler, B.~L.\ 2006, \aap, 458, 69 

\bibitem[1994]{KuijkenDubinski94} Kuijken, K., Dubinski, J.,~1994, MNRAS, 269, 13
\bibitem[1995]{KuijkenDubinski95} Kuijken, K., Dubinski, J.,~1995, MNRAS, 277, 1341


\bibitem[Lin \& Faber 1983]{lin83} Lin, D.~N.~C., \& Faber, 
S.~M.\ 1983, \apjl, 266, L21 

\bibitem[Lisker et al. 2007]{lisker07} Lisker, T., Grebel, 
E.~K., Binggeli, B., \& Glatt, K.\ 2007, \apj, 660, 1186 

\bibitem[Lisker et al. 2006a]{lisker06a} Lisker, T., Glatt, K., 
Westera, P., \& Grebel, E.~K.\ 2006a, \aj, 132, 2432 

\bibitem[Lisker et al. 2006b]{lisker06b} Lisker, T., Grebel, 
E.~K., \& Binggeli, B.\ 2006b, \aj, 132, 497 

\bibitem[Mayer et al. 2001a]{mayer01a} Mayer, L., Governato, F., 
Colpi, M., Moore, B., Quinn, T., Wadsley, J., Stadel, J., \& Lake, G.\ 
2001, \apj, 559, 754 

\bibitem[Mayer et al. 2001b]{mayer01b} Mayer, L., Governato, F., 
Colpi, M., Moore, B., Quinn, T., Wadsley, J., Stadel, J., \& Lake, G.\ 
2001, \apjl, 547, L123 

\bibitem[Mastropietro et al. 2005]{mastropietro05} Mastropietro, C., 
Moore, B., Mayer, L., Debattista, V.~P., Piffaretti, R., \& Stadel, J.\ 
2005, \mnras, 364, 607 

\bibitem[Matkovi{\'c} \& Guzm{\'a}n 2005]{matkovic05} 
Matkovi{\'c}, A., \& Guzm{\'a}n, R.\ 2005, \mnras, 362, 289

\bibitem[MacArthur et al. 2003]{macarthur03} MacArthur, L.~A., 
Courteau, S., \& Holtzman, J.~A.\ 2003, \apj, 582, 689 

\bibitem[M{\'e}ndez-Abreu et 
al. 2008]{mendezabreu08} M{\'e}ndez-Abreu, J., Aguerri, J.~A.~L., Corsini, E.~M., \& Simonneau, E.\ 2008, \aap, 478, 353 

\bibitem[M{\"o}llenhoff 2004]{mollenhoff04} M{\"o}llenhoff, C.\ 2004, \aap, 415, 63  

\bibitem[Moore et al. 1996]{moore96} Moore, B., Katz, N., 
Lake, G., Dressler, A., \& Oemler, A.\ 1996, \nat, 379, 613

\bibitem[Murante et al. 2007]{murante07} Murante, G., Giovalli, 
M., Gerhard, O., Arnaboldi, M., Borgani, S., 
\& Dolag, K.\ 2007, \mnras, 377, 2 
 

\bibitem[Navarro et al. 1997]{navarro97} Navarro, J.~F., Frenk, 
C.~S., \& White, S.~D.~M.\ 1997, \apj, 490, 493 

\bibitem[Nieto et al. 1990]{nieto90} Nieto, J.-L., Davoust, E., Bender, R., \& Prugniel, P.\ 1990, \aap, 230, L17

\bibitem[Nipoti, Londrillo, Ciotti 2003]{nipoti03} Nipoti, C., Londrillo, P., Ciotti, L.,~2003, MNRAS, 344, 748

\bibitem[O'Neill \& Dubinski 2003]{oneil03} O'Neill, J.~K., \& Dubinski, J.\ 2003, \mnras, 346, 251 


\bibitem{}O\~norbe, J., Dom\'{\i}nguez-Tenreiro, R., S\'aiz, A.,
  Artal, H., Serna, A.,~2006, MNRAS, 373, 503

\bibitem[O{\~n}orbe et al. 2007]{onorbe07} O{\~n}orbe, J., 
Dom{\'{\i}}nguez-Tenreiro, R., S{\'a}iz, A., 
\& Serna, A.\ 2007, \mnras, 376, 39 

\bibitem[Pedraz et al. 2002]{pedraz02} Pedraz, S., Gorgas, J., 
Cardiel, N., S{\'a}nchez-Bl{\'a}zquez, P., \& Guzm{\'a}n, R.\ 2002, \mnras, 
332, L59 

\bibitem[Peterson \& Caldwell 1993]{peterson93} Peterson, R.~C., 
\& Caldwell, N.\ 1993, \aj, 105, 1411 

\bibitem[Pfenniger \& Norman 1990]{pfenniger90} Pfenniger, D., \& Norman, C.\ 1990, \apj, 363, 391 

\bibitem[Phillipps et al. 1998]{phillipps98} Phillipps, S., 
Driver, S.~P., Couch, W.~J., \& Smith, R.~M.\ 1998, \apjl, 498, L119 

\bibitem[Pracy et al. 2004]{pracy04} Pracy, M.~B., De Propris, 
R., Driver, S.~P., Couch, W.~J., \& Nulsen, P.~E.~J.\ 2004, \mnras, 352, 
1135 

\bibitem[Quilis et al. 2000]{quilis00} Quilis, V., Moore, B., 
\& Bower, R.\ 2000, Science, 288, 1617 

\bibitem[Roberts 1978]{roberts78} Roberts, M.~S.\ 1978, \aj, 83, 
1026 

\bibitem[Rubin et al. 1985]{rubin85} Rubin, V.~C., Burstein, 
D., Ford, W.~K., Jr., \& Thonnard, N.\ 1985, \apj, 289, 81 

\bibitem[Sakai et al. 2000]{sakai00} Sakai, S., et al.\ 2000, 
\apj, 529, 698 

\bibitem[S{\'a}nchez-Janssen et al. 2008]{sanchezjanssen08} 
S{\'a}nchez-Janssen, R., Aguerri, J.~A.~L., 
\& Mu{\~n}oz-Tu{\~n}{\'o}n, C.\ 2008, \apjl, 679, L77 

\bibitem[Sandage \& Visvanathan 1978]{sandage78} Sandage, A., \& 
Visvanathan, N.\ 1978, \apj, 225, 742 

\bibitem[Sargent et al. 1977]{sargent77} Sargent, W.~L.~W., 
Schechter, P.~L., Boksenberg, A., \& Shortridge, K.\ 1977, \apj, 212, 326 

\bibitem[Schechter 1980]{schechter80} Schechter, P.~L.\ 1980, \aj, 
85, 801 

\bibitem[Schechter \& Gunn 1979]{schechter79} Schechter, P.~L., \& Gunn, J.~E.\ 1979, \apj, 229, 472 

\bibitem[S\`ersic 1968]{sersic68} S\`ersic, J.~L.\ 1968, Cordoba, 
Argentina: Observatorio Astronomico, 1968,  

\bibitem[Shen \& Sellwood 2004]{shen04} Shen, J., \& Sellwood, J.~A.\ 2004, \apj, 604, 614 

\bibitem{} Shu, F. H.,~1969, ApJ, 158, 505

\bibitem[Simien \& de Vaucouleurs 1986]{simien86} Simien, F., \& de Vaucouleurs, G.\ 1986, \apj, 302, 564 

\bibitem[Smith et al. 2004]{smith04} Smith, R.~J., et al.\ 
2004, \aj, 128, 1558 

\bibitem[]{} Springel V.,~2005, MNRAS, 364, 1105

\bibitem[Terlevich et al. 1981]{terlevich81} Terlevich, R., 
Davies, R.~L., Faber, S.~M., \& Burstein, D.\ 1981, \mnras, 196, 381 

\bibitem[Theuns \& Warren 1997]{theus97} Theuns, T., \& 
Warren, S.~J.\ 1997, \mnras, 284, L11 

\bibitem[Tonry 1981]{tonry81} Tonry, J.~L.\ 1981, \apjl, 251, 
L1 

\bibitem[Tormen 1997]{tormen97} Tormen, G. \ 1997, MNRAS, 290, 411

\bibitem[Tremonti et al. 2004]{tremonti04} Tremonti, C.~A., et 
al.\ 2004, \apj, 613, 898 

\bibitem[Tully \& Fisher 1977]{tullyfisher77} Tully, R.~B., \& Fisher, J.~R.\ 1977, \aap, 54, 661 

\bibitem[Trujillo et al. 2002]{trujillo02} Trujillo, I., Asensio 
Ramos, A., Rubi{\~n}o-Mart{\'{\i}}n, J.~A., Graham, A.~W., Aguerri, 
J.~A.~L., Cepa, J., \& Guti{\'e}rrez, C.~M.\ 2002, \mnras, 333, 510 

\bibitem[Trujillo et al. 2001]{trujillo01} Trujillo, I., Aguerri, 
J.~A.~L., Guti{\'e}rrez, C.~M., \& Cepa, J.\ 2001, \aj, 122, 38 

\bibitem[Valenzuela \& Klypin 2003]{valenzuela03} Valenzuela, O., \& Klypin, A.\ 2003, \mnras, 345, 406 

\bibitem[van Zee et al. 2004a]{vanzee04} van Zee, L., Barton, 
E.~J., \& Skillman, E.~D.\ 2004a, \aj, 128, 2797 

\bibitem[van Zee et al. 2004b]{vanzee04b} van Zee, L., Skillman, 
E.~D., \& Haynes, M.~P.\ 2004b, \aj, 128, 121 

\bibitem[Williams et al. 2007]{williams07} Williams, B.~F., et 
al.\ 2007, \apj, 656, 756 

\end{thebibliography}
\end{document}